\documentclass{article}

\usepackage[main, final]{neurips_2026}
\usepackage{wrapfig}
\usepackage[utf8]{inputenc}
\usepackage[T1]{fontenc}
\usepackage{hyperref}
\usepackage{url}
\usepackage{booktabs}
\usepackage{amsmath}
\usepackage{amsfonts}
\usepackage{graphicx}
\usepackage{microtype}
\usepackage{xcolor}

\title{Shared Geometry Is Not Shared Physics:\\
A Layerwise Test of the Platonic Representation Hypothesis in Astronomy}

\author{%
\makebox[\linewidth][c]{%
\begin{tabular}{c}
Kshitij Duraphe$^{1,2}$, Aravind Kannappan$^2$, Dun Li Chan$^{2,3}$, Okiki Famutimi$^2$, Yaswant Sai Ejjagiri$^2$\\[4pt]
Michael J. Smith$^{1,2,4,5}$, John F. Wu$^{6,7}$\\[8pt]
\normalfont $^1$UniverseTBD, $^2$EleutherAI, $^3$INTI International College Penang, Malaysia\\
\normalfont$^4$Center for Astrophysics, Harvard \& Smithsonian, $^5$AstroAI\\
\normalfont $^{6}$Space Telescope Science Institute, $^{7}$Johns Hopkins University
\end{tabular}%
}%
}

\begin{document}

\maketitle

\begin{abstract}
In this paper we investigate whether geometrically aligned astronomical representations are also scientifically interchangeable. 
We analyze every layer's representation from 36 pretrained models using cross-matched optical images, infrared images, and spectra. All 139 analyzed model–-survey comparisons show significant local-neighborhood alignment somewhere in the network after accounting for the search over depth. However, alignment does not generally increase with network depth and can be stronger for the changes between consecutive computational blocks than for the block outputs themselves.
We find that geometry-selected stages under-perform label-selected stages in every image-transfer comparison; moreover, every final HSC--COSMOS--Web model pair is aligned while every redshift transfer has negative $R^2$. Models can therefore recover a similar cross-survey geometry among astronomical objects without recovering a survey-invariant linear encoding of the physical properties tested here.
\end{abstract}

\section{Introduction}

The Platonic Representation Hypothesis \citep{huh2024platonic} (PRH) has gained significant attention in mechanistic interpretability \citep{trott2025theorygeneralizabilityllmmechanistic}, cognitive science \citep{Xu_2025}, and multimodal modeling \citep{koepke2026platoscaveexaminingcrossmodal}. It implies that larger neural networks tend to learn a shared statistical representation of reality, regardless of their training data, architecture, or training method. PRH-like alignment has been observed in chemical language models \citep{kenneth2026chemical} and interatomic potentials \citep{li2026platonic}, but whether it is scientifically useful may depend on whether the aligned views carry redundant or complementary information \citep{tjandrasuwita2025multimodal}.

% For example, chemical language models can organize molecular semantics across their layers \citep{kenneth2026chemical}, while foundation interatomic potentials exhibit PRH-like alignment across independently developed models \citep{li2026platonic}. The natural question is whether this alignment is scientifically useful. Recent work suggests that in the multimodal setting, the answer depends on whether the two views contain redundant or complementary information \citep{tjandrasuwita2025multimodal}.

% \footnote{To reproduce our results, see: \url{https://github.com/ksd3/layerwise-analysis}.}

We examine how PRH-like geometric alignment develops across intermediate and final representations extracted from pretrained foundation models.  \footnote{To reproduce our results, see: \url{https://anonymous.4open.science/r/hidden-topology-3DC3/figures/05_normalized_depth_atlas.py}.} We refer to each extracted activation as a \emph{saved representation stage}. A \emph{computational block} is an architecture-defined processing unit within the model backbone, and its output constitutes one type of saved representation stage. Other stages include embeddings, normalization outputs, pooling stages, projections, and output heads. Our datasets contain observations of astronomical sources together with associated physics labels. 
Neither is a direct \emph{measurement} of the source. Emitted light propagates through space and, for ground-based surveys, Earth’s atmosphere. Telescope optics and detectors then record a distorted and incomplete signal which calibration and reduction transform into a science-ready product; and catalog construction may apply an additional measurement or inference model. These operations are lossy and survey dependent. %John's wording, sounds much cleaner than mine
Therefore, two representations of a crossmatched galaxy are conditioned by different wavelength coverage, resolution, sensitivity, selection, and reduction pipelines, and need not preserve the same physical information. Even tokenization choices can change both the representations learned from galaxy images and their physical prediction performance \citep{sanjaripour2026tokenizer}. We consequently distinguish \emph{geometric alignment}---agreement in neighborhoods, global pairwise structure, or retrieval---from \emph{scientific alignment}, for which the same predictor of a physical quantity must transfer across surveys.

We trace both geometric and scientific alignment across 36 pretrained models using four object-matched datasets. In each dataset, an observation from the Subaru Hyper Suprime-Cam (HSC) \citep{aihara2018hsc} is paired object-by-object with either a DESI spectrum \citep{desi2022instrument} or imaging from the DESI Legacy Imaging Surveys \citep{dey2019legacy}, the James Webb Space Telescope (JWST) \citep{gardner2006jwst}, or COSMOS-Web \citep{casey2023cosmosweb}. We use the crossmatches from \citet{duraphe2025platonic}, excluding SDSS for the domain-validity reason summarized in Appendix~\ref{app:methods}.

We ask three questions: \emph{where} geometric alignment appears, \emph{what} produces it, and \emph{whether} it carries physical information in a transferable form. We find that shared geometry is widespread but moves irregularly through network depth. Most importantly, it neither selects the best stages for physical prediction nor makes final representations scientifically interchangeable. Formal definitions appear in Appendix~\ref{app:alignment-definitions}, and the experimental details appear in Appendix~\ref{app:methods}.

\section{Where Does Alignment Appear?}

For each model-survey comparison, we ask whether the largest alignment \footnote{We measure alignment with mutual $k$-nearest-neighbor overlap (MKNN) \citep{duraphe2025platonic}, Procrustes \citep{schonemann1966procrustes} and CKA \citep{kornblith2019similarity} for global similarity, and MRR \citep{voorhees1999mrr} for exact-object retrieval.} observed across the model’s extracted representations at any stage exceeds what would be expected under an arbitrary correspondence between objects. 
\begin{wrapfigure}{r}{0.63\textwidth}
    \centering
    \includegraphics[width=\linewidth]{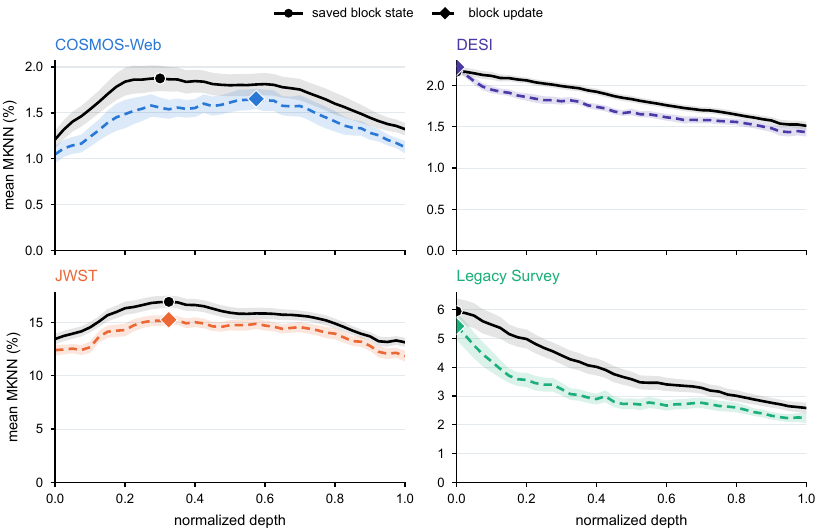}
    \caption{Chance-corrected MKNN (a \emph{depth curve}) across normalized depth for saved block outputs and consecutive differences. For each model with $T$ valid transitions, transition $r$ and its resulting state have depth $d_r=(r-1)/(T-1)$; input embeddings and non-block hooks are excluded. Models are interpolated onto a common depth grid and weighted equally. Solid curves: block states, dashed curves: updates ($\Delta Z_r=Z_{r+1}-Z_r$), bands show standard errors, and markers indicate mean-curve maxima. For image--image comparisons, dashed curves use updates from both views; for DESI, they compare each HSC update with the fixed spectral representation. Vertical scales differ across panels.}
    \label{fig:block-depth}
    %\vspace{-1cm}
\end{wrapfigure}
Because we are asking whether alignment is unusually high at \emph{any} network depth, the null procedure must also allow a high value to occur at any depth by chance. Thus, for each permutation, we shuffle the cross-survey object correspondence, calculate alignment at every network stage, and take the maximum across stages. The resulting permutation maxima form the null distribution against which we compare the observed maximum. All 139 analyzed model–survey comparisons remain significant after correction for multiple tests. Shared geometry is therefore detectable somewhere along network depth in every analyzed comparison. We next ask where it is strongest and whether it remains strong at later depths. 

Alignment does not increase steadily with network depth. One might expect later computations to discard instrumental details and retain a more abstract description of the source, but the mean contrast between the first and final quarters of the saved stages is approximately zero for COSMOS-Web and negative for DESI, JWST, and the Legacy Survey, while the final stage is rarely compatible with the maximum. 
A network can therefore reach strong cross-survey neighborhood alignment at an intermediate stage and then move away from it. Additional depth does not necessarily preserve or strengthen shared geometry. Figure~\ref{fig:block-depth} shows the saved outputs of computational blocks together with the net changes between consecutive outputs.

% Different metrics reveal different depth trends: in most JWST models, exact-object retrieval improves while local-neighborhood agreement declines. Identifying the corresponding object therefore does not imply preserving its neighborhood structure; CKA and Procrustes capture additional global differences (Appendix~\ref{app:metric-depth}). This correspondence between stages is similarly not fixed across views. A same-index comparison assumes that stage $\ell$ in one view represents roughly the same point in the computation as stage $\ell$ in another, which need not hold when the inputs are processed differently. To test this directly, we compare every saved stage in one view with every saved stage in the other. The strongest alignment is usually on or near the same-index diagonal, but exact same-stage matches are substantially more common for the Legacy Survey than for JWST or COSMOS-Web. Thus, stage number provides a useful but imperfect measure of correspondence between the two views (Appendix~\ref{app:cross-layer}).

Different metrics reveal different depth trends: in most JWST models, exact-object retrieval improves while local-neighborhood agreement declines. Identifying the corresponding object therefore does not imply preserving its neighborhood structure; CKA and Procrustes capture additional global differences (Appendix~\ref{app:metric-depth}). Stage correspondence is similarly not fixed across views. A same-index comparison assumes that stage $\ell$ in one view represents roughly the same point in the computation as stage $\ell$ in another, which need not hold when inputs are processed differently. To test this directly, we compare every saved stage in one view with every saved stage in the other. The strongest alignment is usually on or near the same-index diagonal, but exact same-stage matches are substantially more common for the Legacy Survey than for JWST or COSMOS-Web. Thus, stage number provides a useful but imperfect measure of correspondence between the two views (Appendix~\ref{app:cross-layer}).

\section{Where Do Aligned Changes Appear?}

Each saved block output contains the accumulated effects of all preceding computations. Residual architectures motivate viewing depth as iterative refinement \citep{jastrzebski2018residual}. For transformers, successive components are often described as reading from and writing to a shared residual stream \citep{elhage2021mathematical}. Because our models span transformer and convolutional architectures, we use the architecture-neutral term \emph{block-to-block update} for the net difference between consecutive saved block outputs. Our previous depth curves show where alignment is visible, but not necessarily where it is introduced. We therefore examine the change made between consecutive saved stages. For consecutive saved outputs in the same numbered block stack, and only when those outputs have the same representation width, we define the net block-to-block displacement:
$\Delta Z_{m,\ell}^{(s)} = Z_{m,\ell+1}^{(s)} - Z_{m,\ell}^{(s)}$
where $Z_{m,\ell}^{(s)} \in \mathbb{R}^{n \times d}$ contains the representations of the same $n$ objects at stage $\ell$ of model $m$ in survey $s$. The $i^{th}$ row of $\Delta Z_{m,\ell}^{(s)}$ is therefore the change made to object $i$ between the two stages. In residual architectures, it can be interpreted as an aggregate block update, but the exact interpretation depends on where normalization and other operations occur.  We compare the geometry of these differences rather than their Euclidean magnitude. MKNN and retrieval use cosine-normalized representations, while CKA and Procrustes are normalized similarities; uniform growth in activation scale therefore cannot by itself produce the reported alignment curves. Nevertheless, normalization placement can affect what a block-to-block difference represents.
We then measure alignment between $\Delta Z_{m,\ell}^{(a)}$ and $\Delta Z_{m,\ell}^{(b)}$ using the same statistics as for the saved representations. For DESI, only a single spectral representation is available rather than a sequence over spectral-model depth, so each HSC block-to-block update is compared with it.

The update curves are significant in every retained comparison. In a substantial minority of cases, a single block update is more aligned than any valid saved representation. Moreover, the most strongly aligned update typically occurs before the saved representation reaches its maximum, and this ordering holds in nearly every COSMOS-Web curve. Thus, an individual computation can introduce a strongly shared change before its effect becomes most visible in the accumulated representation. Later computations may preserve, reinforce, or obscure that change (Figure~\ref{fig:block-depth}; Appendix~\ref{app:update-timing}). 
 %Downstream compensation following a causal intervention has been described as self-repair or the Hydra effect \citep{mcgrath2023hydra}; however, our analysis does not ablate blocks and therefore does not test that mechanism.

We next test whether the alignment could arise from object count, representation width, or generic high-dimensional geometry. Shuffling object correspondences or replacing representations with independent isotropic Gaussian matrices reduces local alignment to approximately zero reflecting object-specific structure shared across surveys rather than matrix dimensions or random geometry. The effect also survives substantial compression: random 32/64-dimensional projections preserve much of the original local alignment. Removing leading principal components increases local alignment while reducing CKA and Procrustes similarity, suggesting that high-variance directions shape global geometry without determining local neighborhoods. Regressing out magnitude, $g-r$ color, and redshift also leaves the local effect largely intact. Finally, after accounting for representation width, survey interactions, and architecture family, larger models do not show stronger maximum or cumulative alignment; final-stage alignment is, if anything, weaker. Thus, model scale does not appear to progressively amplify shared geometry (Appendices~\ref{app:control-results} and \ref{app:model-properties}).

\section{Does Alignment Encode Shared Physics?}

If geometric alignment reflects shared physical information, then a stage selected \emph{without} using physical labels should still support physical prediction. We test this without allowing the labels to affect the geometric selection. For Legacy Survey color ($g-r$), shape (fitted half-light radius), JWST magnitude (Kron-aperture magnitude) and size (half-light radius), one validation set selects the stage with the best cross-survey retrieval, while a separate validation set selects the stage with the best physical prediction. Both choices are then evaluated on the same held-out objects. The geometry-selected stage has lower $R^2$ in all eight target--direction comparisons, with every family-level interval excluding zero. The reduction in $R^2$ ranges from $0.060$ in the mildest comparison to $3.62$ in the strongest (Figure~\ref{fig:geometry-not-physics}a).

Across models, stronger maximum geometric alignment is associated with stronger maximum prediction of JWST magnitude in both transfer directions (Spearman $\rho=0.48$ and $0.62$). The relationship is absent for JWST size ($\rho=-0.12$ and $0.18$) and weaker for Legacy Survey color and shape ($\rho=0.15$--$0.37$). Thus, even when each model is allowed to reach its best stage under each criterion, maximum geometric alignment is not a general predictor of maximum physical performance (Figure~\ref{fig:max-physics-alignment}). The complete depth maps appear in Appendix~\ref{app:label-depth}.

\begin{figure*}[t]
\centering
\includegraphics[width=\textwidth]{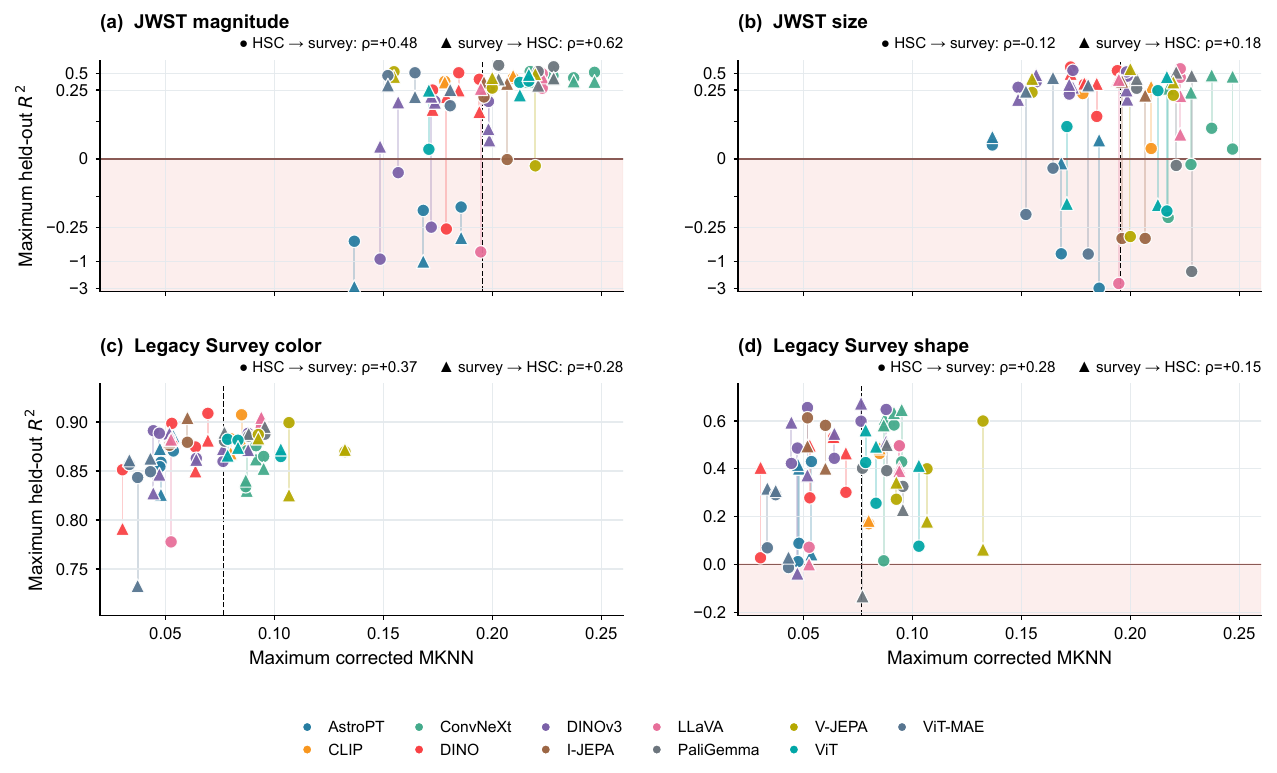}
\caption{Maximum physical prediction versus geometric alignment across 36 models for \textbf{(a)} JWST magnitude, \textbf{(b)} JWST size, \textbf{(c)} Legacy Survey color, and \textbf{(d)} Legacy Survey shape. Circles and triangles indicate HSC-to-survey and reverse transfer, with vertical segments linking directions for each model. Coordinates show maximum valid corrected MKNN and mean held-out linear-probe $R^2$ across representation stages; colors indicate model family, $\rho$ the Spearman correlation, and dashed lines the median maximum MKNN. Shading marks performance below the test-set mean, and JWST panels use a symmetric-log vertical scale. Maxima need not occur at the same stage.}
\label{fig:max-physics-alignment}
\end{figure*}

An explicitly learned rank-64 PCA–-CCA space between HSC images and DESI spectra improves held-out retrieval to mean MRR (0.038), showing that additional object correspondence can be recovered through a learned cross-survey transformation.  However, because this transformation is fitted without a physical target, it does not demonstrate that a physical predictor transfers unchanged between surveys, consistent with domain-adaptation work showing explicit alignment objectives, adapters, and target-domain fine-tuning improve transfer between simulations and observations and across surveys \citep{ciprijanovic2021deepmerge,ciprijanovic2023deepastrouda,walmsley2022practical}. Our experiment instead asks the stricter question of whether independently pretrained representations already provide a survey-invariant linear encoding, without an adapter or target-domain refitting.

We then remove stage selection entirely. On HSC--COSMOS-Web, every final model pair has significant local alignment, and within-survey redshift prediction has mean $R^2=0.503$. However, every cross-survey transfer is negative, with median $R^2=-1.43$ and mean $R^2=-2.98$. Alignment does not predict redshift transfer after accounting for model family, parameter count, representation width, and within-survey performance ($p=0.294$), nor does it predict transfer of color, magnitude, stellar mass, or specific star-formation rate (Figure~\ref{fig:geometry-not-physics}b).

\begin{figure}[h]
\centering
\includegraphics[width=\textwidth]{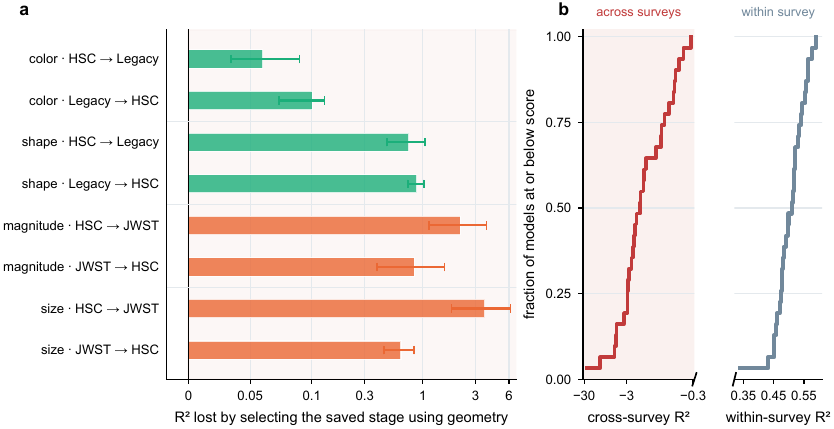}
\caption{\textbf{(a)} Test-set $R^2$ loss from selecting a saved representation stage by geometry rather than physical labels. Bars show family-equal mean losses with 95\% family-bootstrap intervals; all exclude zero. \textbf{(b)} Empirical CDFs of within- and cross-survey redshift $R^2$ for 31 aligned final HSC--COSMOS-Web model pairs. All within-survey scores are positive and all cross-survey scores are negative. A broken horizontal axis shows both distributions at readable scales.}
\label{fig:geometry-not-physics}
\end{figure}

We observe that neighborhood agreement constrains relations among objects not the direction that predicts a particular physical quantity. Two representations can therefore organize the same galaxies similarly while requiring different physical predictors.

\section{Discussion and Conclusion}

Our results support a restricted interpretation of the PRH. Models applied to different observations of the same astronomical objects recover shared geometries, but not a survey-invariant linear representation of physical information. Alignment is a useful label-free diagnostic of whether two views organize objects similarly and where that agreement emerges in the model. However, it is insufficient for choosing representations for scientific inference using the same linear probe across surveys; this requires either target-specific validation or a model trained explicitly for cross-survey transfer.

% Future work? Explicit modeling of differing data/survey systematics. Exploration of mech interp representations e.g. using crosscoders/cross-layer transcoders. Investigation of representational vs functional similarities.

Future work should identify the sources and functional consequences of shared geometry. Survey-conditioned objectives could model differences in bandpasses, PSFs, noise, calibration, and selection \citep{ciprijanovic2023deepastrouda}. Sparse crosscoders could separate shared from survey-specific transformer features \citep{lindsey2024sparse}, while cross-layer transcoders and attribution graphs could test their computational roles \citep{ameisen2025circuit,chatzoudis2026crosslayer}. Causal interventions should then test whether perturbing shared features produces matched downstream effects across views, since representational similarity may capture decodable rather than functionally used information \citep{hayne2024representation,braun2025dissociation}.
\bibliographystyle{plainnat}
\bibliography{references}

\appendix

\section{Methods and Experimental Details}
\label{app:methods}

\subsection{Definitions of alignment}
\label{app:alignment-definitions}

We use the same dataset as the Platonic Universe paper
\citep{duraphe2025platonic} and the foundation models described
below. For a model $m$ and layer $\ell$, let
\[
Z_{m,\ell}^{(s)}\in\mathbb{R}^{n\times d_{m,\ell,s}}
\]
denote the representations of $n$ crossmatched objects observed in survey
$s$. The rows of $Z_{m,\ell}^{(a)}$ and $Z_{m,\ell}^{(b)}$ therefore correspond
to the same physical objects, although their observed appearances and
representation widths may differ.

We define \emph{saved representation stage} as any intermediate or terminal representation, \emph{computational block} as a block in the model architecture, and \emph{layer} only when referring to an architecture's native indexing (i.e. attention layer). A \emph{block update} is the difference between two compatible consecutive block outputs.  Thus, a saved stage may be an input embedding, normalization output, pooling stage, projection, or output head, rather than a conventional layer.

\subsection{Datasets and models}

HSC \citep{aihara2018hsc} is the reference dataset in every survey pairing. We pair it with DESI \citep{desi2016design} spectra, Legacy Survey \citep{dey2019legacy} images, JWST \citep{gardner2006jwst} images, and COSMOS-Web \citep{casey2023cosmosweb} images. We exclude SDSS because prior validation of these data found that the DESI-trained SpecFormer encoder remains out of domain for SDSS even after wavelength resampling, so its SDSS representations are not scientifically interpretable \citep{duraphe2025platonic}.

Table~\ref{tab:datasets} separates the number of rows in each released representation product from the number used in the analysis. For DESI, the Legacy Survey, and JWST, we retain rows only when both cross-survey object identifiers are present and unique; if an identifier appears more than once, every occurrence is removed rather than selecting one arbitrarily. The retained rows are then placed in a stable object-identifier order shared by every model. The historical JWST image-quality filter is reapplied before this identifier filtering. COSMOS-Web layerwise representations already use their released row order and require no further reduction.

By the \emph{final COSMOS-Web product}, we mean a collection of terminal representations for 45,000 crossmatched HSC--JWST sources. For each of 31 models, it contains one HSC embedding matrix and one JWST embedding matrix, with row $i$ in both matrices representing the same source. The corresponding physical catalog consists of the first 45,000 rows of
\href{https://huggingface.co/datasets/Ashodkh/cosmosweb-hsc-jwst-high-snr-pil2}{\texttt{Ashodkh/cosmosweb-hsc-jwst-high-snr-pil2}}, retained in their released order. From this catalog we use spectroscopic redshift, stellar mass, specific star-formation rate (sSFR), HSC $g-r$ color, and HSC $r$-band magnitude. This product is distinct from the 56,322-object layerwise COSMOS-Web product: the two have different rows and object orderings, and we do not join them by row number or use one to make claims about the other.

\begin{table*}[t]
\centering
\small
\renewcommand{\arraystretch}{1.10}
\setlength{\tabcolsep}{4pt}

\caption{Crossmatched datasets used in the experiments. ``Released rows'' is
the size of the representation product; ``analyzed rows'' is the size after
the deterministic filtering described in the text. ``Models'' counts complete
layerwise curves in the first four rows and terminal-representation pairs in
the final row.}
\label{tab:datasets}

\vspace{2pt}

\resizebox{0.96\textwidth}{!}{%
\begin{tabular}{lrrrl}
\toprule
HSC paired with
& Released rows
& Analyzed rows
& Models
& Primary use \\
\midrule
DESI spectra
& 20,465 & 20,321 & 36
& Geometry and common-space retrieval \\
Legacy Survey images
& 101,725 & 101,619 & 36
& Color and shape transfer \\
JWST images
& 1,496 & 929 & 36
& Magnitude and size transfer \\
COSMOS-Web images
& 56,322 & 56,322 & 31
& Layerwise geometry \\
COSMOS-Web final product
& 45,000 & 45,000 & 31
& Five-target final transfer \\
\bottomrule
\end{tabular}%
}

\end{table*}

The 36 models belong to 11 broad families: AstroPT, CLIP, ConvNeXt, DINO,
DINOv3, I-JEPA, LLaVA, PaliGemma, ViT, ViT-MAE, and V-JEPA
\citep{smith2024astropt,radford2021clip,liu2022convnext,caron2021dino,
simeoni2025dinov3,assran2023ijepa,liu2024llava,steiner2024paligemma2,
dosovitskiy2021vit,he2022mae,bardes2024vjepa}. We retain every available
model rather than selecting models according to their results.
Table~\ref{tab:model-basket} gives the exact variants; the released artifact
additionally enumerates representation files, training categories, parameter
counts, widths, and saved-stage names.

\vspace{0.4em}

\begin{table*}[t]
\centering
\small
\renewcommand{\arraystretch}{1.12}
\setlength{\tabcolsep}{5pt}

\caption{The complete model basket. Names follow the released representation
files.}
\label{tab:model-basket}

\vspace{2pt}

\begin{tabular}{@{}lp{0.78\textwidth}@{}}
\toprule
Family & Model variants \\
\midrule
AstroPT &
\texttt{astropt\_015M}, \texttt{astropt\_095M},
\texttt{astropt\_850M} \\
CLIP &
\texttt{clip\_base}, \texttt{clip\_large} \\
ConvNeXt &
\texttt{convnext\_nano}, \texttt{convnext\_tiny},
\texttt{convnext\_base}, \texttt{convnext\_large} \\
DINO &
\texttt{dino\_small}, \texttt{dino\_base},
\texttt{dino\_large}, \texttt{dino\_giant} \\
DINOv3 &
\texttt{dinov3\_vits16}, \texttt{dinov3\_vits16plus},
\texttt{dinov3\_vitb16}, \texttt{dinov3\_vitl16},
\texttt{dinov3\_vith16plus}, \texttt{dinov3\_vit7b16} \\
I-JEPA &
\texttt{ijepa\_huge}, \texttt{ijepa\_giant} \\
LLaVA &
\texttt{llava\_15\_7b}, \texttt{llava\_15\_13b},
\texttt{llava\_ov\_7b} \\
PaliGemma &
\texttt{paligemma\_3b}, \texttt{paligemma\_10b},
\texttt{paligemma\_28b} \\
ViT &
\texttt{vit\_base}, \texttt{vit\_large}, \texttt{vit\_huge} \\
ViT-MAE &
\texttt{vit-mae\_base}, \texttt{vit-mae\_large},
\texttt{vit-mae\_huge} \\
V-JEPA &
\texttt{vjepa\_large}, \texttt{vjepa\_huge},
\texttt{vjepa\_giant} \\
\bottomrule
\end{tabular}

\end{table*}

\subsection{Experiments}
\label{app:experiments}

We organize the analysis into nine experiments. These experiments test where
alignment appears, how it is produced, and also whether it depends on model properties,
and whether it transfers to physical prediction.

\begin{table}[t]
\centering
\small
\renewcommand{\arraystretch}{1.18}
\setlength{\tabcolsep}{3pt}
\setlength{\abovecaptionskip}{0.5em}

\caption{List of experiments.}
\label{tab:experiments}

\vspace{2pt}

\begin{tabular}{@{}p{0.07\linewidth}p{0.86\linewidth}@{}}
\toprule
ID & Question \\
\midrule
E1 & Is local alignment significant anywhere along each complete depth curve? \\
E2 & Does alignment increase from early to late saved stages? \\
E3 & Does label-free geometric selection identify scientifically useful saved stages? \\
E4 & Does an explicitly fitted common space improve image--spectrum retrieval? \\
E5 & Which consecutive block updates write the shared geometry? \\
E6 & Do model scale, architecture, or training objective explain alignment? \\
E7 & Do the two image views reach their strongest alignment at the same saved-stage index? \\
E8 & Does terminal-representation alignment predict cross-survey performance on physical labels? \\
E9 & Which representation components and possible nuisance variables explain the alignment? \\
\bottomrule
\end{tabular}

\end{table}

\subsection{Geometry}

For two representations $X$ and $Y$ containing the same $n$ objects, we normalize each row and compute exact cosine nearest neighbors. If $N_k^X(i)$ denotes the neighbor set of object $i$, the raw overlap is
\[
O_k(X,Y)=\frac{1}{n}\sum_{i=1}^{n}\frac{|N_k^X(i)\cap N_k^Y(i)|}{k}.
\]
The expected overlap of two random sets is $p_0=k/(n-1)$, giving the corrected statistic
\[
A_k(X,Y)=\frac{O_k(X,Y)-p_0}{1-p_0}.
\]
We use $k=20$ as the primary value and compute $k\in\{5,10,50,100\}$ as sensitivity analyses retained in the released artifact. Exact top-100 neighbors are computed once and reused for all values of $k$. Self-matches are removed by row identity.

Let $\widetilde X=X-\mathbf{1}\bar{x}^{\mathsf T}$ and $\widetilde Y=Y-\mathbf{1}\bar{y}^{\mathsf T}$ denote feature-centered representations. Centered linear CKA is
\[
\operatorname{CKA}(X,Y)=
\frac{\lVert \widetilde X^{\mathsf T}\widetilde Y\rVert_F^2}
{\lVert \widetilde X^{\mathsf T}\widetilde X\rVert_F
 \lVert \widetilde Y^{\mathsf T}\widetilde Y\rVert_F},
\]
and normalized Procrustes similarity is
\[
\operatorname{Proc}(X,Y)=
\frac{\lVert \widetilde X^{\mathsf T}\widetilde Y\rVert_*}
{\lVert\widetilde X\rVert_F\lVert\widetilde Y\rVert_F},
\]
where $\lVert\cdot\rVert_*$ is the nuclear norm. Both are calculated in float64. For exact paired-object retrieval, let $r_i^{X\rightarrow Y}$ be the cosine-similarity rank of row $i$ of $Y$ when row $i$ of $X$ is the query. We report the bidirectional statistic
\[
\operatorname{MRR}(X,Y)=\frac{1}{2n}\sum_{i=1}^{n}
\left(\frac{1}{r_i^{X\rightarrow Y}}+
      \frac{1}{r_i^{Y\rightarrow X}}\right),
\]
using midranks for exact ties. Constant, nonfinite, or tie-dominated representations are marked unavailable rather than assigned a chance value. Image--image datasets use same-stage curves and complete cross-stage matrices. Image--spectrum datasets compare every HSC saved stage with the fixed released spectral representation.

\subsection{Permutation tests and uncertainty}

For each originally evaluated depth curve, we generate 9,999 permutations of the cross-survey object correspondence. We then save the maximum alignment across the entire curve as our statistic. The resulting $p$-value therefore includes the stage search. 
%The 139 retained curves are assessed

The 139 non-SDSS curves analyzed in this paper are assessed using the Holm correction over all 175 originally evaluated curves, which is more conservative than correcting only the retained set. We use 2,000 paired-object bootstrap draws for curve uncertainty and resample whole model families for cross-model conclusions. Secondary physical targets use false-discovery-rate correction. In the large Legacy Survey pairing, each permutation evaluates one fixed, seeded sample of 8,192 query rows to keep the null calculation tractable; the observed curve and every bootstrap draw use all 101,619 analyzed objects.

The simultaneous curve band is constructed from shared bootstrap row draws, so dependence among saved stages is retained. We define the simultaneous peak set as every stage whose upper simultaneous bound is at least the largest lower simultaneous bound on the curve. The final stage is called compatible with the maximum only when it belongs to this set.

For a released curve with $L$ ordered saved stages, stage $j\in\{0,\ldots,L-1\}$ has normalized saved-stage depth $j/(L-1)$. The early and final quarters used in E2 are depths at most $0.25$ and at least $0.75$, respectively; their mean difference is the primary directional depth summary. Figures restricted to computational blocks remove input embeddings and non-block hooks before defining their block-depth coordinates, as stated in their captions.

For the smaller image and spectral datasets, five fixed outer folds are constructed with seed 42 from deterministic object partitions. Each fold assigns approximately 60\% of rows to fitting, 10\% to geometry selection, 10\% to label selection, and 20\% to final testing. These four roles are disjoint. The final COSMOS-Web experiment instead sorts objects by right ascension and divides them into ten contiguous slabs. In each of five folds, two slabs are tested, their neighboring slabs are discarded as spatial buffers, two distant slabs select regularization, and the remaining slabs are fitted. Every object appears in a test set exactly once.

\subsection{Scientific transfer}

The layerwise image-transfer targets are defined directly from the released catalogs. Legacy Survey color is $-2.5\log_{10}(\mathrm{FLUX\_G}/\mathrm{FLUX\_R})$ and Legacy Survey shape is $\log_{10}(\mathrm{SHAPE\_R})$, with nonpositive radii excluded. JWST size is $\log_{10}(\mathrm{flux\_radius})$, again requiring a positive radius, and JWST magnitude is \texttt{mag\_auto}. In the final COSMOS-Web product, redshift is \texttt{specz} restricted to $0<z<20$, with rows having \texttt{Confidence\_level}$<0$ excluded; color is HSC model magnitude $g-r$, magnitude is HSC model magnitude $r$, stellar mass is \texttt{lp\_mass}, and specific star-formation rate is \texttt{lp\_ssfr}. Missing or invalid targets are omitted using the same finite rows within each fit, selection, and evaluation role.

E3 fits linear ridge probes and compares two saved-stage choices. Only same-width source and target representations are eligible, because the fitted source transformation and coefficients are applied to the target without refitting. Source-view standardization is fitted on the fit rows and transferred unchanged. Ridge strength is selected from $\{0.01,0.1,1,10,100\}$ using the label-validation rows. The geometry-selected stage is the earliest stage within one standard error of the best bidirectional paired retrieval on the geometry-validation rows, without using physical labels. The label-selected stage maximizes validation $R^2$ on a different set of rows. Both are evaluated on the same held-out test objects. We report
\[
\mathrm{regret}=R^2_{\mathrm{geometry\ selected}}-R^2_{\mathrm{label\ selected}},
\]
so a negative value means that geometry selected the less useful stage. We first average folds and checkpoints within each architecture family, then give every family equal weight. The intervals in Figure~\ref{fig:geometry-not-physics}a are percentile intervals from 2,000 bootstrap resamples of those family means; they therefore do not treat related checkpoints or the five folds as independent observations.

Table~\ref{tab:e3-stage-distance} describes how far apart the two selectors land. For each model and transfer task, we first average the signed normalized-stage difference across the five folds, then summarize the resulting 36 model-level values. Positive values mean that the label-selected stage is deeper; negative values mean that it is shallower. The broad interquartile ranges show that the disagreement is not a fixed displacement shared by all models.

\begin{table*}[t]
\centering
\small
\renewcommand{\arraystretch}{1.10}
\setlength{\tabcolsep}{3.5pt}

\caption{Distribution of the normalized saved-stage difference
$\Delta=s_{\mathrm{label}}-s_{\mathrm{geometry}}$. Each row contains 36
model-level values after averaging the five folds. $Q_{25}$ and $Q_{75}$ are
the first and third quartiles.}
\label{tab:e3-stage-distance}

\vspace{2pt}

\resizebox{0.94\textwidth}{!}{%
\begin{tabular}{lllrrrrr}
\toprule
Survey & Target & Transfer & $Q_{25}$ & Median $\Delta$
& $Q_{75}$ & Median $|\Delta|$ & Range \\
\midrule
Legacy Survey & Color & HSC$\rightarrow$Legacy
& $-0.380$ & $-0.042$ & $0.117$ & $0.237$ & $[-0.912,\ 0.400]$ \\
Legacy Survey & Color & Legacy$\rightarrow$HSC
& $-0.295$ & $0.127$ & $0.322$ & $0.322$ & $[-0.877,\ 0.731]$ \\
Legacy Survey & Shape & HSC$\rightarrow$Legacy
& $0.029$ & $0.160$ & $0.599$ & $0.240$ & $[-0.857,\ 0.848]$ \\
Legacy Survey & Shape & Legacy$\rightarrow$HSC
& $-0.180$ & $0.286$ & $0.614$ & $0.523$ & $[-0.882,\ 0.857]$ \\
JWST & Magnitude & HSC$\rightarrow$JWST
& $-0.300$ & $-0.140$ & $0.034$ & $0.237$ & $[-0.594,\ 0.800]$ \\
JWST & Magnitude & JWST$\rightarrow$HSC
& $-0.267$ & $-0.098$ & $0.281$ & $0.269$ & $[-0.800,\ 0.690]$ \\
JWST & Size & HSC$\rightarrow$JWST
& $-0.132$ & $0.019$ & $0.205$ & $0.178$ & $[-0.671,\ 0.720]$ \\
JWST & Size & JWST$\rightarrow$HSC
& $-0.200$ & $-0.042$ & $0.065$ & $0.167$ & $[-0.629,\ 0.483]$ \\
\bottomrule
\end{tabular}%
}

\end{table*}

\begin{table*}[t]
\centering
\scriptsize
\renewcommand{\arraystretch}{1.10}
\setlength{\tabcolsep}{3.6pt}

\caption{Mean held-out $R^2$ lost by selecting a saved representation stage
using geometry rather than physical labels, reported separately by model
family. H, L, and J denote HSC, Legacy Survey, and JWST.}
\label{tab:e3-family-loss}

\vspace{2pt}

\resizebox{0.98\textwidth}{!}{%
\begin{tabular}{lrrrrrrrr}
\toprule
& \multicolumn{4}{c}{Legacy Survey}
& \multicolumn{4}{c}{JWST} \\
\cmidrule(lr){2-5}
\cmidrule(lr){6-9}
Family
& Color H$\to$L
& Color L$\to$H
& Shape H$\to$L
& Shape L$\to$H
& Magnitude H$\to$J
& Magnitude J$\to$H
& Size H$\to$J
& Size J$\to$H \\
\midrule
AstroPT
& 0.084 & 0.040 & 0.375 & 0.692
& 4.665 & 4.120 & 3.618 & 0.598 \\
CLIP
& 0.041 & 0.055 & 1.001 & 0.931
& 1.400 & 0.365 & 1.828 & 0.843 \\
ConvNeXt
& 0.015 & 0.190 & 1.605 & 1.099
& 0.931 & 0.433 & 2.818 & 0.445 \\
DINO
& 0.045 & 0.093 & 0.745 & 1.003
& 1.361 & 0.402 & 0.674 & 0.262 \\
DINOv3
& 0.097 & 0.088 & 0.269 & 0.408
& 1.679 & 0.152 & 0.265 & 0.275 \\
I-JEPA
& 0.016 & 0.142 & 1.746 & 0.937
& 8.385 & 1.154 & 13.714 & 1.327 \\
LLaVA
& 0.007 & 0.029 & 0.465 & 0.519
& 1.187 & 0.400 & 3.118 & 0.947 \\
PaliGemma
& 0.122 & 0.073 & 0.248 & 1.313
& 0.548 & 0.157 & 2.481 & 0.223 \\
ViT
& 0.029 & 0.164 & 0.859 & 1.062
& 1.445 & 0.546 & 0.871 & 0.722 \\
ViT-MAE
& 0.159 & 0.145 & 0.356 & 0.910
& 0.762 & 0.573 & 2.982 & 0.380 \\
V-JEPA
& 0.048 & 0.093 & 0.537 & 0.852
& 1.601 & 0.975 & 7.450 & 0.993 \\
\bottomrule
\end{tabular}%
}

\end{table*}

E4 standardizes the HSC and DESI representations, reduces each to rank 64 with PCA, and fits regularized CCA using only the fit rows. The CCA ridge is selected from $\{0.001,0.01,0.1,1\}$, and the saved stage and ridge strength are chosen jointly by bidirectional retrieval on the geometry-selection rows. Every standardizer, PCA basis, and CCA transformation is fitted without the held-out test rows. DESI contributes held-out retrieval results but has no physical label in this experiment, so E4 is treated as a geometric correspondence test rather than a scientific-transfer test.

E8 uses the final 45,000-object COSMOS-Web product and the spatial folds described above. For each physical target, we fit within-survey and cross-survey ridge probes in both directions, pooling residual sums of squares and target moments across the five held-out slabs before calculating one $R^2$ per model and direction. The primary regression asks whether local alignment predicts cross-survey redshift $R^2$ after accounting for parameter count, representation width, within-survey performance, and architecture family. Alignment significance is Holm-corrected across the 31 terminal model pairs. Regression significance is calculated by permuting alignment within family, and uncertainty is obtained by resampling whole families.

\subsection{Descriptive depth of physical-label probes}
\label{app:label-depth}

Neural networks often fit simple, low-frequency, or broadly shared structure before more irregular patterns \citep{arpit2017memorization,rahaman2019spectral,kalimeris2019complexity}. Therefore, we expect ``closer-to-instrument'' labels to be easier to probe. Specifically, apparent magnitude and color are close to quantities recorded by the instrument; for example, $g-r$ follows directly from fluxes measured in two passbands. Galaxy size and morphology instead depend on the spatial distribution of light, while redshift requires identifying or inferring spectral or photometric features. Stellar mass and specific star-formation rate are further removed from the measurements and require additional physical inference. If network depth corresponds even approximately to increasing levels of abstraction, instrument-proximal quantities might therefore become accessible at shallower stages than quantities that require several stages of inference.

Depth is not a direct proxy for the order in which a model learned its features. A model can also expose a useful statistic at one saved stage without producing a representation that is globally aligned with another observational view. We therefore treat this hierarchy as a descriptive expectation, not as a prediction that alignment must increase monotonically with depth.

The physical labels provide a separate way to ask whether this progression corresponds to increasing scientific complexity. In the Legacy Survey experiments, the label-selected stage for observed $g-r$ color is shallower than the stage selected for fitted shape in $69\%$ of matched model, fold, and transfer-direction comparisons. This is consistent with the idea that an instrument-proximal statistic can become accessible before a fitted structural quantity. The ordering is not universal: for JWST, magnitude is selected earlier than size in $47.5\%$ of comparisons and later in $47.8\%$. These counts are descriptive rather than independent statistical tests, since folds and models within a family are dependent.

For this label-depth comparison, we convert each E3 label-selected stage to normalized saved-stage depth and compare the two targets within the same model, fold, and transfer direction. In Legacy Survey transfer, median depth is $0.238$ for color and $0.786$ for shape in the HSC$\rightarrow$Legacy direction; the corresponding values are $0.529$ and $0.724$ in the reverse direction. In JWST transfer, median depth is $0.360$ for magnitude and $0.583$ for size in the HSC$\rightarrow$JWST direction, but $0.540$ and $0.429$ in the reverse direction. These comparisons reuse folds and models and are therefore presented as descriptive evidence rather than a separate inferential experiment.

\begin{figure*}[p]
\centering
\includegraphics[width=\textwidth]{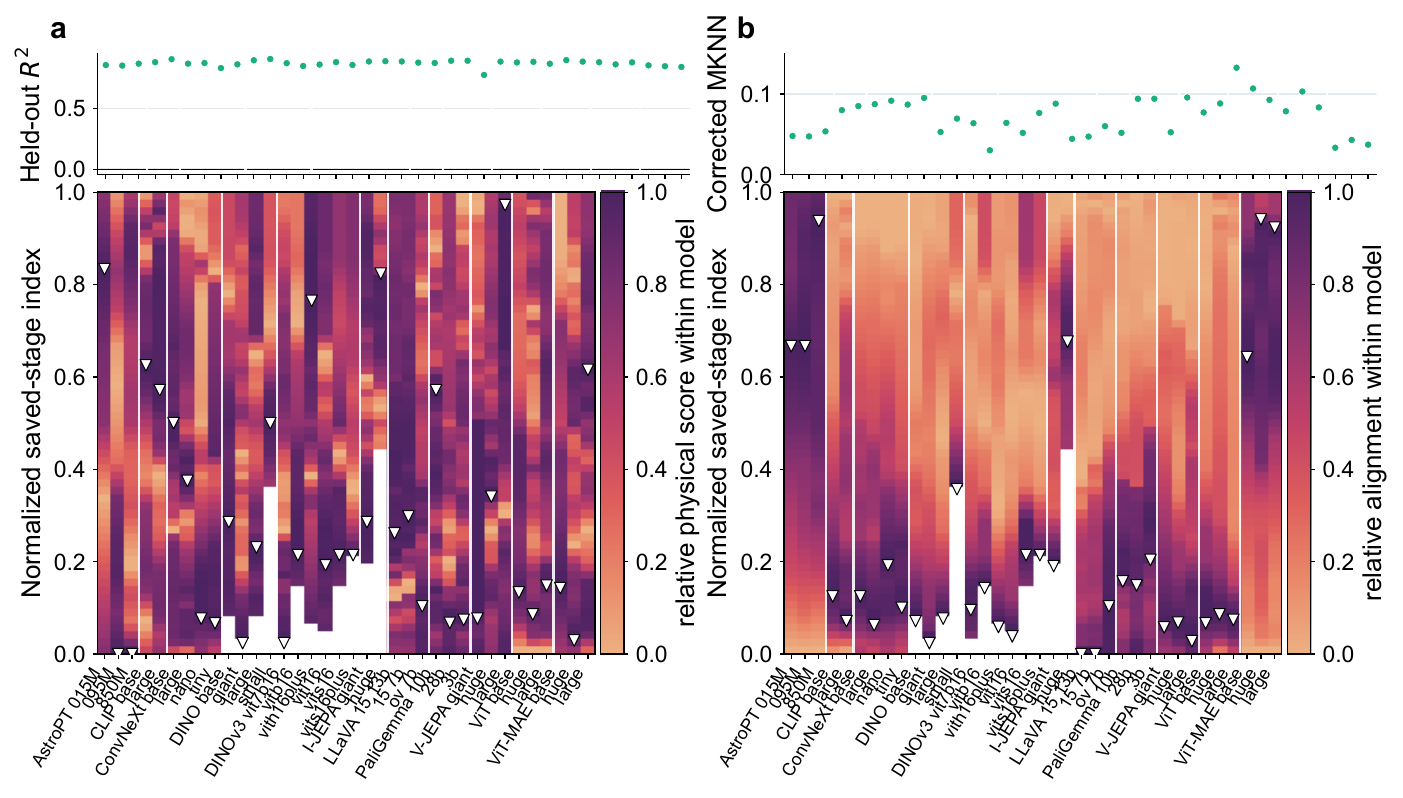}
\medskip
\includegraphics[width=\textwidth]{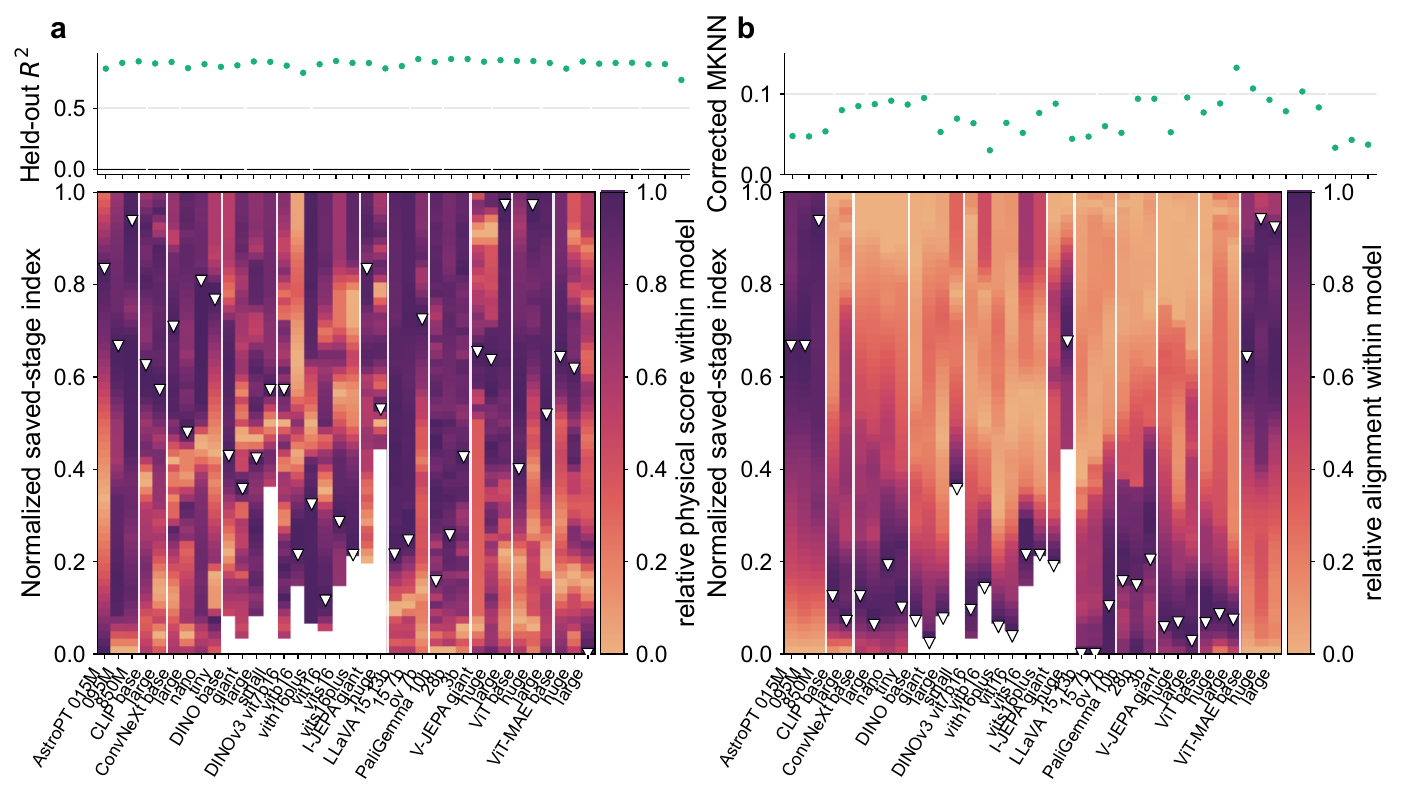}
\caption{Saved-stage Legacy Survey color prediction and geometric alignment for all 36 models. The upper panel transfers a color predictor from HSC to the Legacy Survey; the lower panel transfers in the reverse direction. In each panel, the left map shows held-out $R^2$ and the right map shows corrected MKNN. Each model column is scaled from its own minimum to maximum so that depth structure remains visible; the scatter plots above retain the actual model-level maxima, and white triangles locate them along the normalized saved-stage index.}
\label{fig:label-depth-legacy-color}
\end{figure*}

\begin{figure*}[p]
\centering
\includegraphics[width=\textwidth]{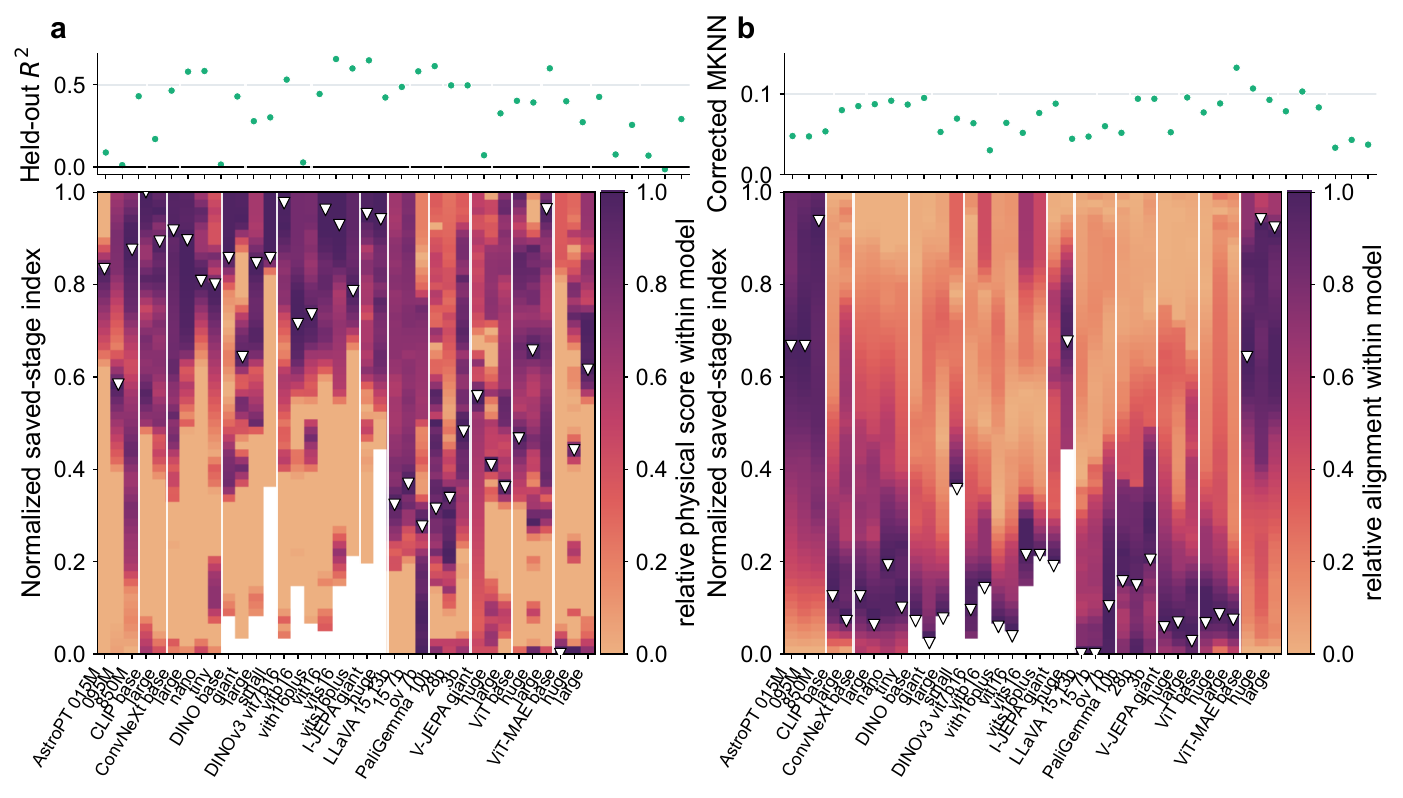}
\medskip
\includegraphics[width=\textwidth]{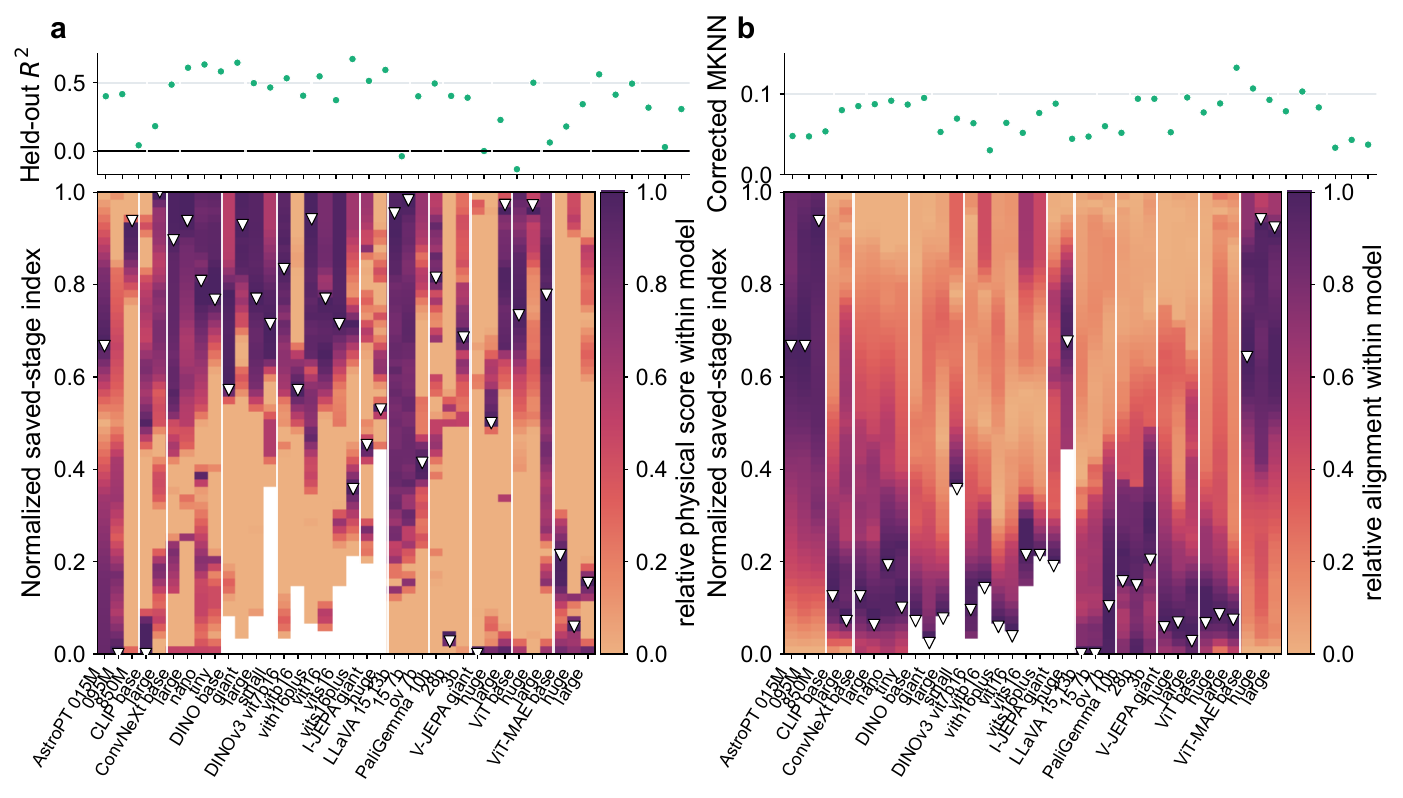}
\caption{Saved-stage Legacy Survey shape prediction and geometric alignment for all 36 models. The upper panel transfers from HSC to the Legacy Survey and the lower panel transfers in the reverse direction. Plotting conventions follow Figure~\ref{fig:label-depth-legacy-color}.}
\label{fig:label-depth-legacy-shape}
\end{figure*}

\begin{figure*}[p]
\centering
\includegraphics[width=\textwidth]{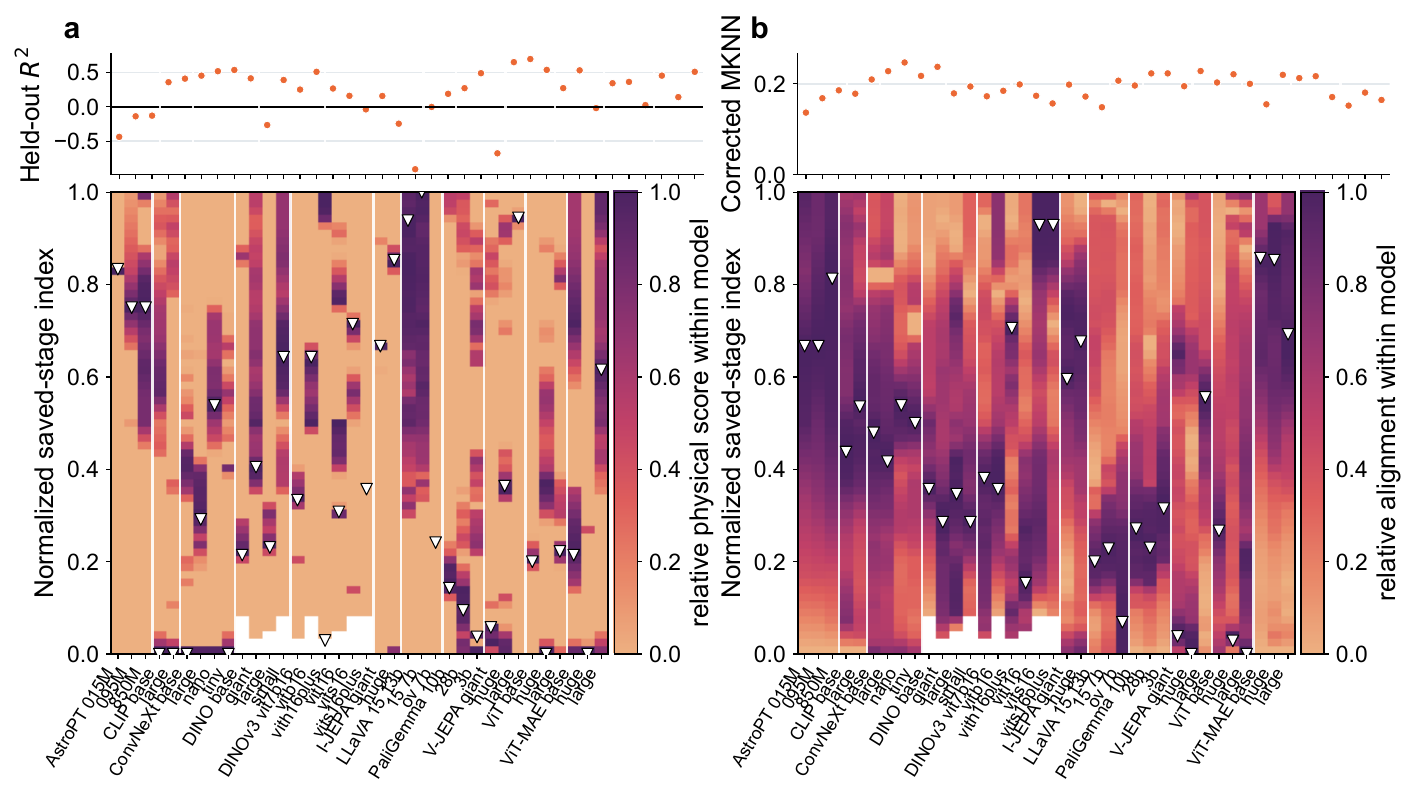}
\medskip
\includegraphics[width=\textwidth]{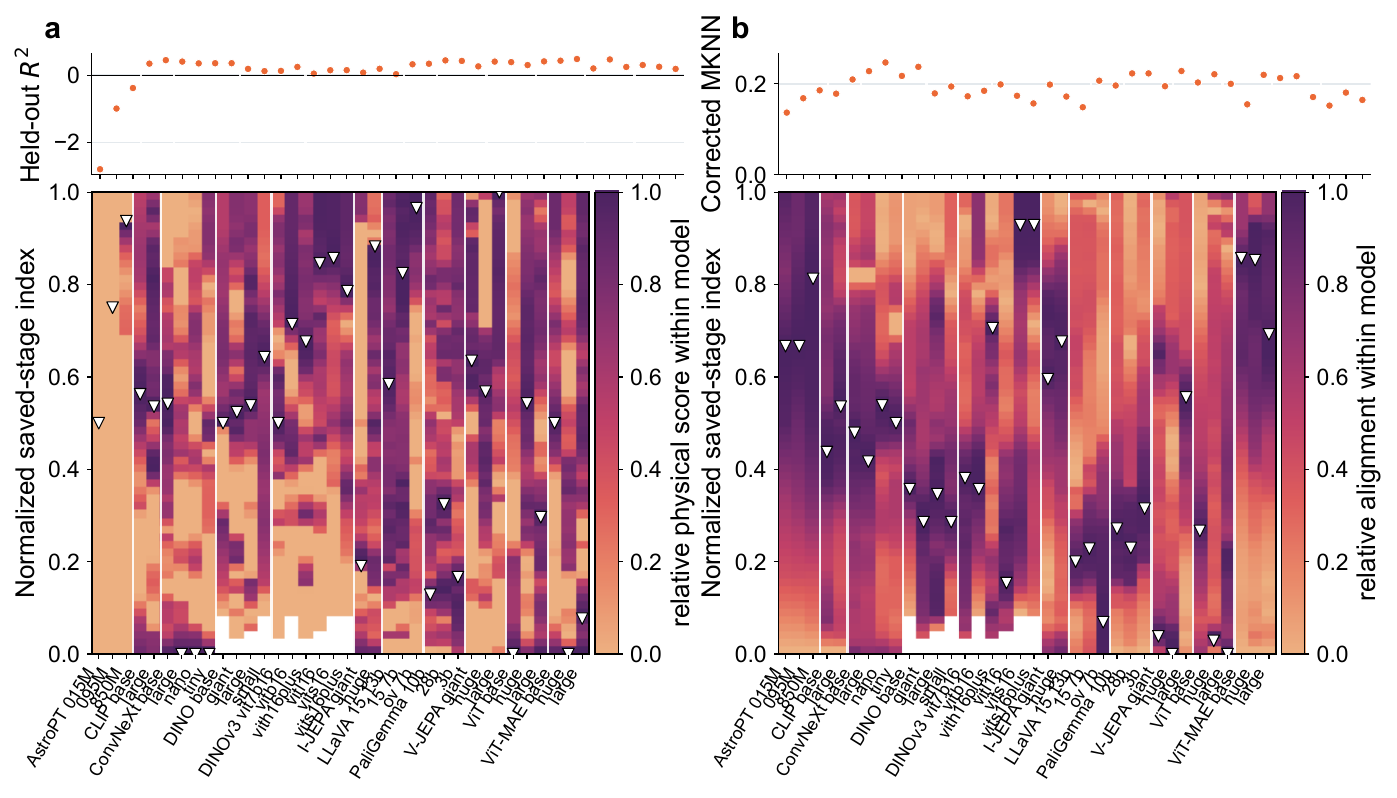}
\caption{Saved-stage JWST magnitude prediction and geometric alignment for all 36 models. The upper panel transfers from HSC to JWST and the lower panel transfers in the reverse direction. Plotting conventions follow Figure~\ref{fig:label-depth-legacy-color}.}
\label{fig:label-depth-jwst-magnitude}
\end{figure*}

\begin{figure*}[p]
\centering
\includegraphics[width=\textwidth]{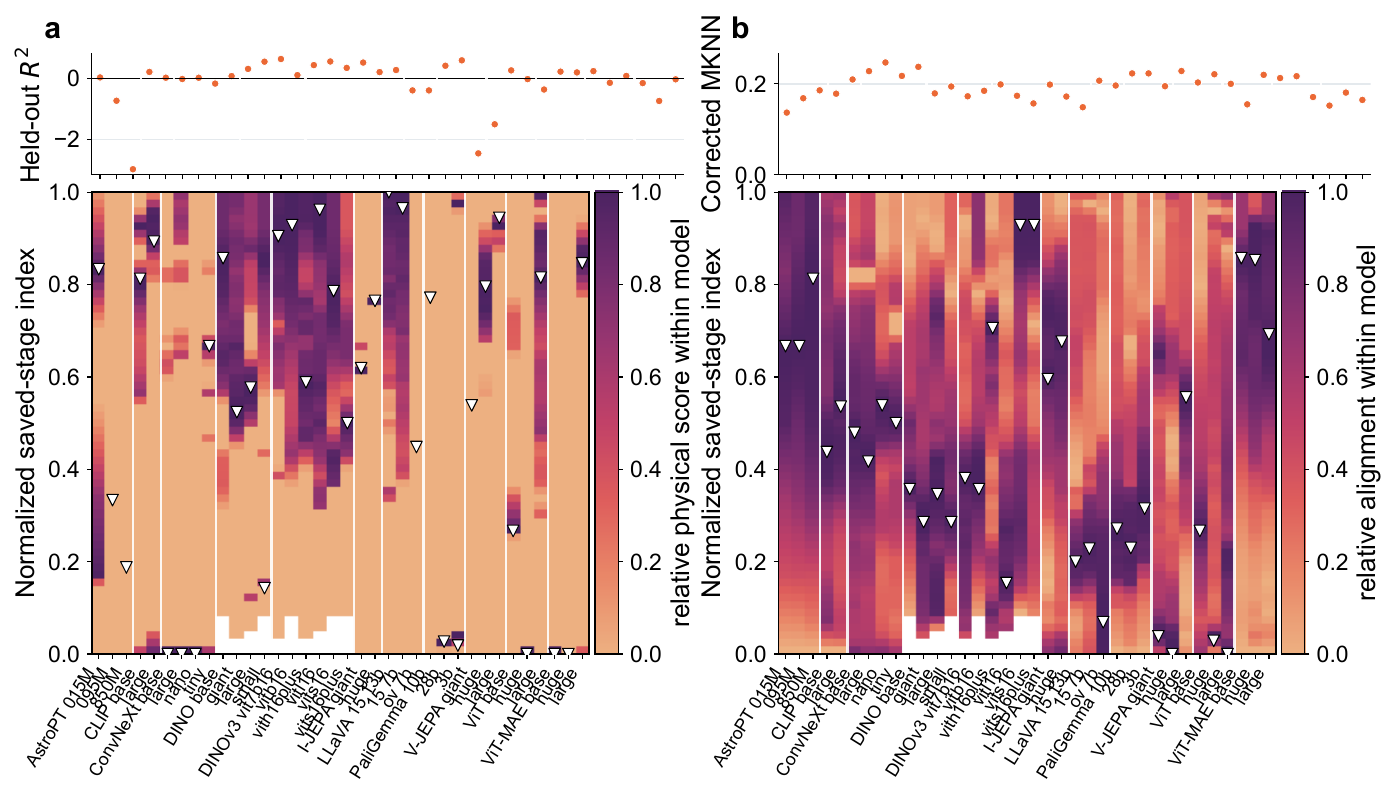}
\medskip
\includegraphics[width=\textwidth]{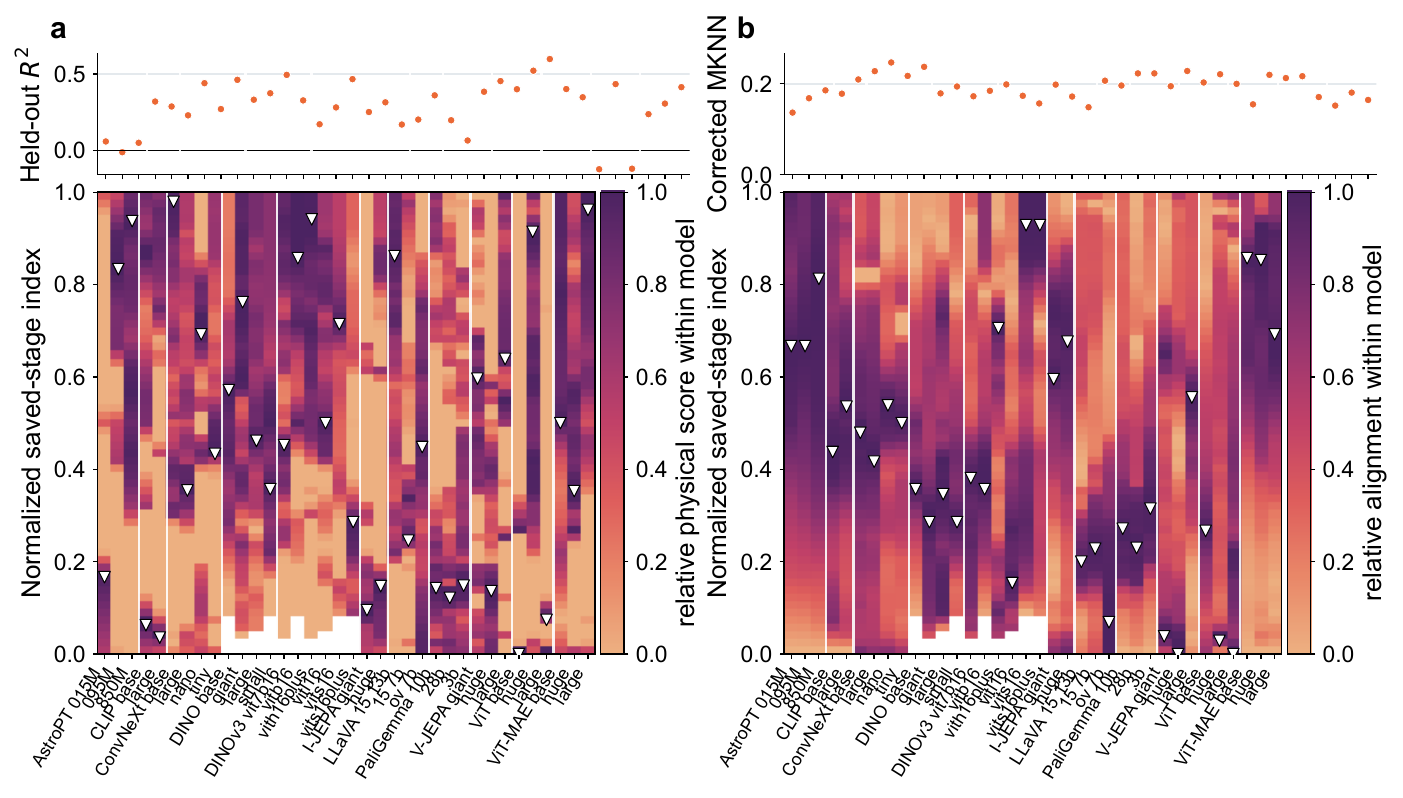}
\caption{Saved-stage JWST size prediction and geometric alignment for all 36 models. The upper panel transfers from HSC to JWST and the lower panel transfers in the reverse direction. Plotting conventions follow Figure~\ref{fig:label-depth-legacy-color}.}
\label{fig:label-depth-jwst-size}
\end{figure*}

Figures~\ref{fig:label-depth-legacy-color}--\ref{fig:label-depth-jwst-size} show the complete saved-stage results behind this comparison. At each saved stage, the physical score is the mean held-out $R^2$ across five folds. The maximum over this test curve is descriptive; the disjoint validation procedure described above remains the basis of the stage-selection result in the main text. White triangles mark the depth of each maximum, and unavailable or tie-dominated representations are omitted from both sides of the comparison.

\subsection{Block updates, model properties, and controls}

E5 is restricted to consecutive computational-block outputs with compatible widths and row order. For image--image comparisons, we align the two block updates defined in the main text. DESI has one fixed released spectral representation rather than a spectral depth curve, so each HSC block update is instead compared with that fixed representation. We compare the peak update alignment with the strongest valid E1 saved-representation alignment; unavailable or tie-dominated input stages are not treated as valid maxima.

E6 regresses alignment summaries on parameter count, representation width, architecture family, and broad training objective. Parameter-count significance is evaluated through within-family permutation and family-level bootstrap intervals. Leave-one-family-out prediction is reported separately to distinguish description of the observed families from generalization to an unseen family.

E9 applies matched transformations to the final COSMOS-Web representations. The negative controls shuffle object correspondence, shuffle both views, or replace both representations with independent isotropic Gaussian matrices matched only in row and column shape. Other controls project to 32 or 64 dimensions, remove leading principal components using fit rows, jointly regress magnitude, HSC $g-r$ color, and redshift from both representations, or add isotropic noise to identical representations. Transformations are fitted without using evaluation rows. Local overlap, CKA, Procrustes similarity, and MRR are recomputed after every transformation. Table~\ref{tab:e9-controls} reports the principal controls on a shared scale.

\begin{table*}[t]
\centering
\scriptsize
\renewcommand{\arraystretch}{1.10}
\setlength{\tabcolsep}{3.5pt}

\caption{Alignment retained after each control, expressed as a percentage of
the corresponding unmodified mean metric. PC removal is compared with raw
held-out representations, and catalog regression with raw complete-case
held-out representations; the other controls use the full-sample raw
representations. Values above $100\%$ indicate that the intervention
increases the metric.}
\label{tab:e9-controls}

\vspace{2pt}

\resizebox{0.96\textwidth}{!}{%
\begin{tabular}{llrrrr}
\toprule
Control & Matched reference & MKNN & CKA & Procrustes & MRR \\
\midrule
Shuffle correspondence
& Raw
& $0.027\%$ & $0.075\%$ & $7.325\%$ & $10.116\%$ \\

Matched Gaussian data
& Raw
& $0.022\%$ & $10.133\%$ & $33.896\%$ & $10.008\%$ \\

Random projection to 32 dimensions
& Raw
& $50.987\%$ & $80.656\%$ & $89.378\%$ & $47.879\%$ \\

Random projection to 64 dimensions
& Raw
& $70.577\%$ & $91.299\%$ & $93.243\%$ & $64.616\%$ \\

Remove first five principal components
& Raw held-out
& $128.367\%$ & $37.341\%$ & $67.705\%$ & $478.818\%$ \\

Regress magnitude, color, and redshift
& Raw complete-case held-out
& $96.887\%$ & $70.858\%$ & $87.771\%$ & $315.826\%$ \\
\bottomrule
\end{tabular}%
}

\end{table*}

\subsection{Handling large output heads and approximate reductions}

Several vision--language representations contain 152,128 or 257,216 dimensions. Directly constructing their feature cross-products is infeasible. For bias-free logits $Z=HW^\top$, we accumulate $G=W^\top W$ and construct a factor $L$ satisfying $LL^\top=G$. The reduced representation $HL$ obeys
\[
(HL)(HL)^\top=ZZ^\top,
\]
so it preserves the row Gram matrix used by cosine neighbors, linear CKA, normalized Procrustes similarity, and same-width retrieval. Relative validation errors are below $1.6\times10^{-5}$ for the affected models. Randomized GPU PCA in E4 and E9 is deterministic given its seed.

\section{Additional Results}

\subsection{Metric-dependent depth dynamics}
\label{app:metric-depth}

% Different measures of alignment also give different pictures of this progression. In JWST, exact-object retrieval improves toward later saved stages for $83\%$ of the models, while mean MKNN, CKA, and Procrustes similarity decrease. COSMOS-Web shows a related separation: MRR increases in $81\%$ of models, compared with $58\%$ for local MKNN, $35\%$ for CKA, and $29\%$ for Procrustes similarity. These are not contradictory results. They measure different properties of a representation: global pairwise structure, local neighborhoods, and exact-object retrieval need not evolve together. Thus, even when two representations become more similar according to one definition, this does not imply that they are becoming more interchangeable in every sense. Figure~\ref{fig:normalized-depth-atlas} shows the corresponding block-state and block-update curves for CKA, Procrustes similarity, and MRR.

\begin{figure*}[p]
\centering
\includegraphics[width=\textwidth]{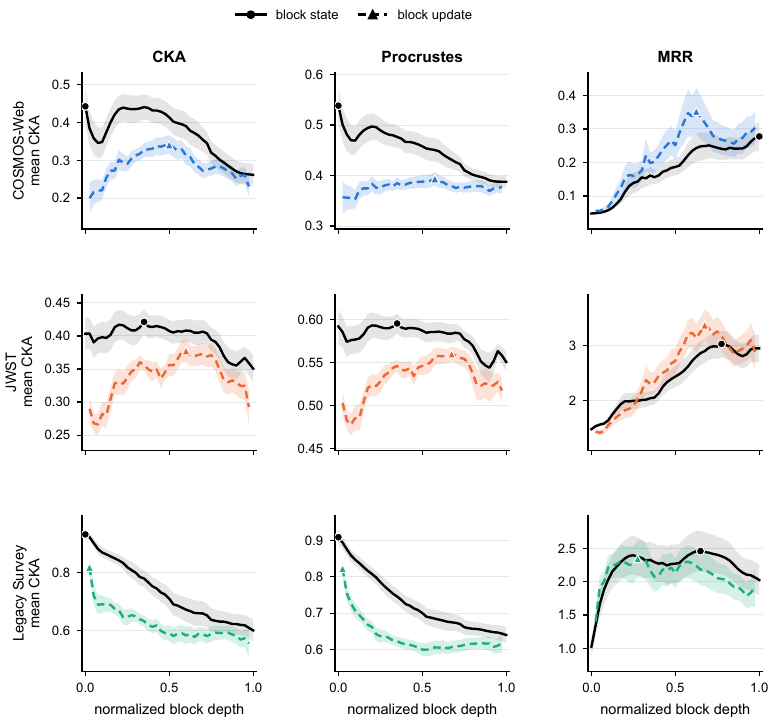}
\caption{Block-state and block-update alignment over normalized computational depth for the three image--image comparisons. Input embeddings, normalization-only hooks, pooling outputs, projections, and output heads are excluded so that depth zero consistently denotes the first computational block. For $B$ retained block states, state $j$ is placed at $j/(B-1)$ and update $j\!\rightarrow\!j+1$ at its midpoint $(j+1/2)/(B-1)$. Solid dark curves show block states; dashed colored curves show consecutive block updates, with circles and triangles marking their respective mean-curve maxima. Each curve averages models with valid first and final block representations after placing them on a common block-depth grid; bands show one standard error across models. Figure~\ref{fig:block-depth} instead assigns an update and the state it produces to the same transition coordinate so that their trajectories can be compared directly. Vertical scales differ between panels.}
\label{fig:normalized-depth-atlas}
\end{figure*}

Depth trends depend on the metric. In JWST, MRR increases from the first to the final quarter in $83\%$ of models, whereas mean MKNN, CKA, and Procrustes similarity decrease. In COSMOS-Web, the corresponding fractions are $81\%$ for MRR, $58\%$ for MKNN, $35\%$ for CKA, and $29\%$ for Procrustes similarity. MRR measures the rank of the paired object, MKNN compares local neighborhoods, and CKA and Procrustes summarize more global structure; the measures need not move together. Figure~\ref{fig:normalized-depth-atlas} shows the block-state and block-update curves for all three secondary metrics.

\subsection{Cross-layer alignment}
\label{app:cross-layer}

% There is a further complication in comparing depth across views. A same-index comparison implicitly assumes that saved stage $\ell$ in one view corresponds to saved stage $\ell$ in the other. E7 removes this assumption by evaluating every saved stage against every other saved stage. Among 103 model--survey comparisons, 63 have their maximum on the same-index diagonal and 83 are within one stage of it. The agreement is particularly strong for the Legacy Survey, where 34 of 36 maxima lie on the diagonal. It is less synchronous for JWST (14 of 36) and COSMOS-Web (15 of 31). A saved-stage index specifies a position in the released representation sequence, but it does not guarantee that two observational views have reached the same computational stage. Figure~\ref{fig:cross-layer-atlas} shows the computational-block subset of these cross-stage comparisons.

\begin{figure*}[p]
\centering
\includegraphics[width=\textwidth]{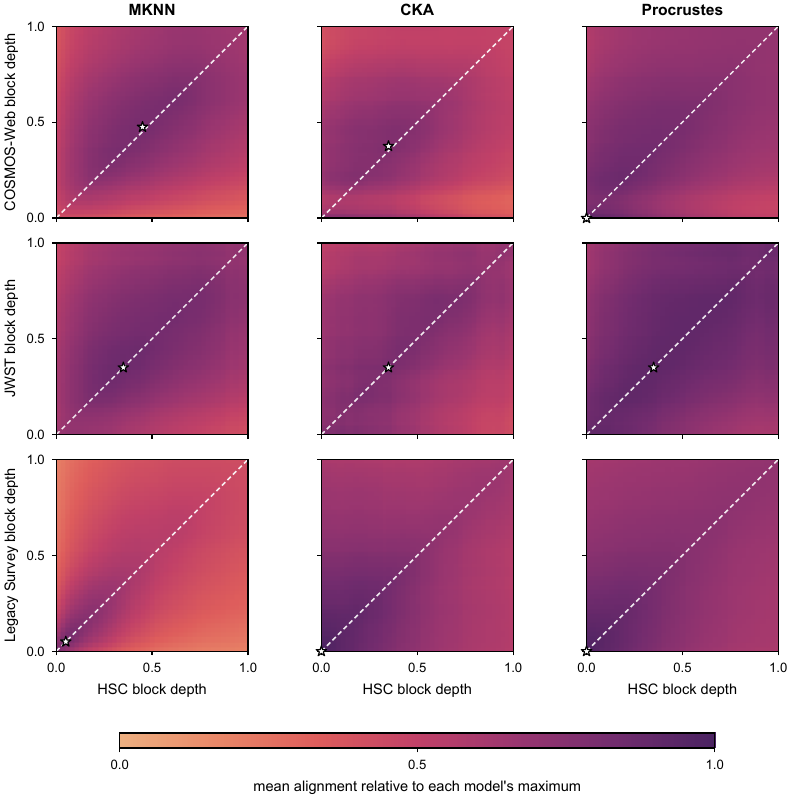}
\caption{Cross-block alignment for the three image--image comparisons. Input embeddings and non-block hooks are excluded, and every remaining HSC block is compared with every block in the paired survey. Each model's map is placed on a common normalized block-depth grid and divided by its own maximum before models are averaged, preventing models with larger metric baselines from dominating the timing pattern. Rows denote surveys and columns denote alignment measures. Dashed lines mark equal block depth and stars mark the maxima of the averaged relative maps.}
\label{fig:cross-layer-atlas}
\end{figure*}

E7 compares every saved stage in one view with every saved stage in the other. Of 103 model--survey maps, 63 reach their maximum on the same-index diagonal and 83 within one stage of it. Same-index maxima occur in 34 of 36 Legacy Survey models, 14 of 36 JWST models, and 15 of 31 COSMOS-Web models. Saved-stage index is therefore a useful but survey-dependent proxy for computational correspondence. Figure~\ref{fig:cross-layer-atlas} shows the computational-block subset of these maps.

% \begin{figure*}[p]
% \centering
% \includegraphics[width=\textwidth]{final_figures/04_metric_and_layer_timing.pdf}
% \caption{Alignment does not follow one saved-stage progression. \textbf{(a)} Fraction of models in which each metric increases from the first to the final quarter of saved stages. DESI MRR is available for nine same-width models; every other displayed survey--metric cell uses all models with a valid metric. \textbf{(b)} Distance from the same-index diagonal of each maximum in the complete cross-stage map; numbers are model counts.}
% \label{fig:metric-layer-timing}
% \end{figure*}

\subsection{Timing of block updates}
\label{app:update-timing}

E5 uses the difference between consecutive block outputs to separate the change made at a block from the representation already present at its input. Alignment in these updates locates shared changes along the network, but does not show that a particular block causes a later peak in representation alignment.

All 139 update curves remain significant after Holm correction. In 46 comparisons, the maximum update alignment is higher than the alignment of every valid saved representation. The median ratio between the update maximum and the saved-representation maximum is $1.01$ for COSMOS-Web, $0.98$ for DESI, $0.97$ for JWST, and $0.87$ for the Legacy Survey.

The update maximum occurs before the saved-representation maximum in $71\%$ of curves and $90\%$ of COSMOS-Web curves. In $83\%$ of comparisons, stronger update alignment is followed by an increase in representation alignment, although only 13 individual associations remain significant after correction. We therefore use the update curves to locate aligned changes, not to make causal claims about individual blocks. Figures~\ref{fig:block-depth} and~\ref{fig:normalized-depth-atlas} show the update and representation curves under the two depth conventions.

\subsection{Model-property results}
\label{app:model-properties}

\paragraph{Does scaling compound the effect?}
We also tested whether the geometric--physical relationship becomes stronger as models grow. It does not appear to compound systematically with parameter count. After controlling for representation width, survey, parameter-count-by-survey interactions, and architecture family, the coefficient of log parameter count was small for maximum alignment ($\beta=-0.0028$, permutation $p=0.0875$), area under the depth curve ($\beta=-0.0025$, $p=0.0773$), and maximum block-update alignment ($\beta=0.0028$, $p=0.0654$). The coefficients were negative for final-stage alignment ($\beta=-0.0069$, $p=0.0027$) and for the final-minus-early alignment difference ($\beta=-0.0094$, $p=0.0014$). Thus, larger models do not simply amplify shared geometry throughout the network; if anything, their alignment is relatively less concentrated at the final stage. These results argue against interpreting scale as a general mechanism that compounds the observed geometric effects.

Table~\ref{tab:e6-properties} reports the SDSS-excluded coefficient summaries. The parameter-count effect is considered supported only when both the within-family permutation test and the family-bootstrap interval agree; none of the six outcomes satisfies both conditions. The training contrasts are family-equal descriptive comparisons and remain confounded with architecture and pretraining data.

\begin{table*}[t]
\centering
\small
\renewcommand{\arraystretch}{1.15}
\setlength{\tabcolsep}{5pt}

\caption{Model-property summaries after excluding SDSS. \textbf{(a)}
Coefficients from regressions containing log parameter count, log
representation width, survey indicators, parameter-count-by-survey
interactions, and architecture-family indicators. Inference was prespecified
for the log-parameter coefficient; width is shown descriptively.
\textbf{(b)} Family-equal differences in maximum MKNN for broad training
categories relative to all other families.}
\label{tab:e6-properties}

\vspace{2pt}

\textbf{\small (a) Scale and width coefficients}

\vspace{0.3em}

\begin{tabular}{lrrrr}
\toprule
Outcome
& $\beta_{\log\mathrm{parameters}}$
& Permutation $p$
& Family 95\% interval
& $\beta_{\log\mathrm{width}}$ \\
\midrule
Maximum alignment
& $-0.0028$ & $0.0875$ & $[-0.0322,\ 0.0123]$ & $0.0234$ \\
Final-stage alignment
& $-0.0069$ & $0.0027$ & $[-0.0441,\ 0.0279]$ & $0.0215$ \\
Area under depth curve
& $-0.0025$ & $0.0773$ & $[-0.0217,\ 0.0102]$ & $0.0201$ \\
Peak saved-stage index
& $0.0336$ & $0.6146$ & $[-0.2838,\ 0.2223]$ & $-0.3170$ \\
Final minus early alignment
& $-0.0094$ & $0.0014$ & $[-0.0529,\ 0.0185]$ & $0.0189$ \\
Maximum update alignment
& $0.0028$ & $0.0654$ & $[-0.0281,\ 0.0215]$ & $0.0046$ \\
\bottomrule
\end{tabular}

\vspace{0.9em}

\textbf{\small (b) Descriptive training-category contrasts}

\vspace{0.3em}

\begin{tabular}{lrr}
\toprule
Category
& Difference in maximum alignment
& Family 95\% interval \\
\midrule
Astronomy self-supervision
& $-0.0214$ & $[-0.0273,\ -0.0145]$ \\
Joint embedding
& $0.0035$ & $[-0.0055,\ 0.0134]$ \\
Masked reconstruction
& $-0.0237$ & $[-0.0295,\ -0.0167]$ \\
Self-distillation
& $-0.0014$ & $[-0.0148,\ 0.0133]$ \\
Supervised images
& $0.0098$ & $[0.0002,\ 0.0205]$ \\
Vision--language training
& $0.0098$ & $[-0.0013,\ 0.0198]$ \\
Astronomy-specific training data
& $-0.0214$ & $[-0.0270,\ -0.0148]$ \\
\bottomrule
\end{tabular}

\end{table*}
\subsection{Common-space retrieval}
\label{app:common-space}

One possible explanation is that HSC and DESI encode compatible information in different forms. We test this explicitly by learning a rank-64 linear common space between HSC images and DESI spectra using PCA followed by regularized CCA (E4). The common space produces nonzero held-out retrieval, with mean MRR of $0.038$, demonstrating that a linear transformation can increase correspondence between the representations. Because this experiment does not include a physical target, it establishes geometric correspondence rather than scientific transfer. The distinction between those two claims is tested directly by the labeled image experiments and the final COSMOS-Web analysis.

\begin{figure*}[t]
\centering
\includegraphics[width=0.78\textwidth]{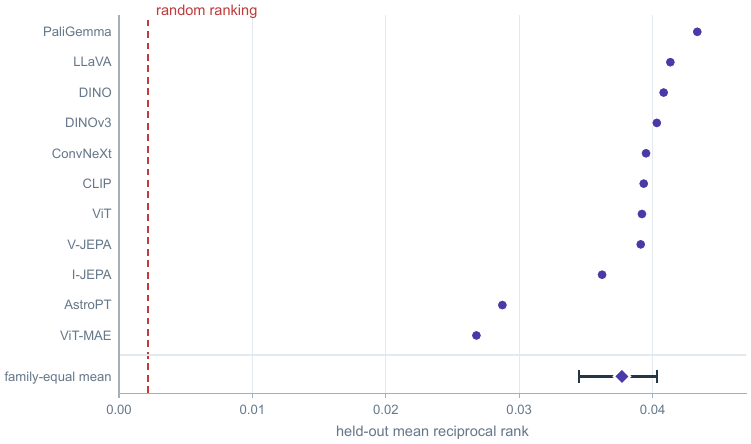}
\caption{Held-out HSC--DESI retrieval after fitting the rank-64 PCA--CCA common space. Points show the 11 architecture-family means; the diamond and interval show their family-equal mean and 95\% family-bootstrap interval. The dashed line marks the fold-averaged random-ranking expectation.}
\label{fig:common-space-retrieval}
\end{figure*}

% \begin{figure*}[t]
% \centering
% \includegraphics[width=0.78\textwidth]{final_figures/common_space_retrieval.pdf}
% \caption{Held-out HSC--DESI retrieval after fitting the rank-64 PCA--CCA common space. Points show means for the 11 architecture families; the diamond and interval show the family-equal mean and its 95\% family-bootstrap interval. Random-ranking expectation is calculated from the five fold-specific test-set sizes for the 20,321 analyzed objects. Mean MRR is $0.0377$ (95\% interval $[0.0345,0.0403]$), compared with random-ranking expectation $0.002187$, a factor of $17.24$. An exact two-sided family sign-flip test gives $p=0.00098$.}
% \label{fig:common-space-retrieval}
% \end{figure*}

\subsection{Interpretation of physical transfer}

% The negative result is therefore not that aligned representations contain no physical information. The within-survey probes show that they can contain substantial information about physical quantities. Rather, the information that is useful for a scientific prediction does not appear to be preserved in a form that transfers unchanged across these observational views. Two representations can preserve similar relationships among objects while supporting different linear probes of a particular physical quantity. Geometric alignment and scientific interchangeability are therefore distinct propositions. For the astronomical setting considered here, substantial alignment is evidence that two models have learned related structure from their inputs, but it is not by itself evidence that they have learned a shared scientific representation.

% Table~\ref{tab:e8-targets} gives the adjusted alignment coefficient for every final COSMOS-Web physical target. The redshift test is primary; false-discovery-rate correction is applied across the four secondary targets. No family-bootstrap interval excludes zero and no target is significant after the applicable correction.

In E8, ridge coefficients fitted in one survey are applied without refitting to representations from the other survey, and corrected MKNN is used to predict the resulting cross-survey $R^2$. Table~\ref{tab:e8-targets} reports the MKNN coefficient for each final COSMOS-Web target after adjustment for parameter count, representation width, within-survey performance, and model family. We report the redshift test without a multiple-testing adjustment and control the false discovery rate across the four $p$-values for color, magnitude, stellar mass, and specific star-formation rate. Every family-bootstrap interval includes zero, and none of the five tests reaches its applicable significance threshold.

\begin{table*}[t]
\centering
\small
\renewcommand{\arraystretch}{1.15}
\setlength{\tabcolsep}{5pt}
\caption{Coefficient of corrected MKNN when predicting cross-survey
physical-transfer $R^2$. Every regression also contains parameter count,
representation width, within-survey performance, and architecture-family
indicators. The redshift test is primary; $q$ is the Benjamini--Hochberg value
across the four secondary targets.}
\vspace{2pt}
\begin{tabular}{lrrrr}
\toprule
Physical target
& Alignment coefficient
& Family 95\% interval
& Permutation $p$
& Secondary $q$ \\
\midrule
HSC $g-r$ color
& $-95.33$ & $[-860.36,\ 906.19]$ & $0.6225$ & $0.6868$ \\
Magnitude
& $-172.92$ & $[-603.43,\ 201.39]$ & $0.3622$ & $0.6868$ \\
Stellar mass
& $-118.73$ & $[-320.25,\ 36.50]$ & $0.0369$ & $0.1476$ \\
Redshift (primary)
& $-349.24$ & $[-1624.34,\ 113.66]$ & $0.2943$ & --- \\
Specific star-formation rate
& $-90.29$ & $[-395.73,\ 69.96]$ & $0.6868$ & $0.6868$ \\
\bottomrule
\end{tabular}

\label{tab:e8-targets}
\end{table*}

\vspace{0.6em}

Finally, Table~\ref{tab:leave-family-out} reports whether these model-property
relationships remain useful when an entire architecture family is omitted from
fitting. The E6 scale--width model is stable for alignment magnitude summaries
but only marginally better than the mean for timing summaries. The E8 redshift
model beats the training-mean baseline for six of ten omitted families, with
nearly identical mean squared error overall.

\vspace{0.3em}

\begin{table*}[t]
\centering
\small
\renewcommand{\arraystretch}{1.15}
\setlength{\tabcolsep}{6pt}

\caption{Leave-one-family-out stability. ``Families improved'' counts
held-out families for which the fitted model has lower squared error than the
training-mean baseline. The MSE ratio is the mean fitted-model MSE divided by
the mean baseline MSE; values below one favor the fitted model. E8
leave-family-out evaluation was prespecified for redshift.}
\label{tab:leave-family-out}

\vspace{2pt}

\begin{tabular}{llrr}
\toprule
Analysis & Outcome & Families improved & Mean MSE ratio \\
\midrule
E6 & Maximum alignment
& $11/11$ & $0.088$ \\
E6 & Final-stage alignment
& $11/11$ & $0.086$ \\
E6 & Area under depth curve
& $11/11$ & $0.064$ \\
E6 & Maximum update alignment
& $11/11$ & $0.079$ \\
E6 & Final minus early alignment
& $6/11$ & $0.930$ \\
E6 & Peak saved-stage index
& $6/11$ & $0.939$ \\
E8 & Cross-survey redshift $R^2$
& $6/10$ & $0.976$ \\
\bottomrule
\end{tabular}

\end{table*}

\subsection{Control results}
\label{app:control-results}

A different possibility is that the observed alignment is produced by a trivial or very low-dimensional component of the representation. We test this using the final COSMOS-Web representations (E9). Table~\ref{tab:e9-controls} summarizes the E9 controls on the final COSMOS-Web representations. Shuffled correspondence and independent shape-matched Gaussian matrices reduce corrected MKNN to approximately zero. Random projections to 32 or 64 dimensions retain approximately $51\%$ and $71\%$ of the original MKNN, while removing the first five principal components raises MKNN to $128\%$ of its matched baseline and lowers CKA and Procrustes similarity. A linear adjustment for magnitude, HSC $g-r$ color, and redshift retains approximately $97\%$ of MKNN. Small isotropic perturbations of otherwise identical representations leave CKA near one but reduce MKNN to approximately $0.70$. The controls rule out the number of objects, the representation dimensions, and the fitted linear nuisance model as sufficient explanations, while the contrasting metric responses show that local neighborhoods and global representational similarity capture different structure.

%Shuffling the cross-survey correspondence or replacing the representations with independent shape-matched isotropic Gaussian data reduces local alignment to approximately zero, showing that the effect is not produced by the metric alone. At the same time, random projections to 32 and 64 dimensions retain approximately $51\%$ and $71\%$ of the original local alignment. The shared neighborhood structure is therefore substantially compressible, but it is not confined to a small number of dominant components. Removing the leading principal components actually increases local alignment while decreasing CKA and Procrustes similarity. Jointly regressing magnitude, HSC $g-r$ color, and redshift from both views likewise does not remove the local effect.

%These controls also clarify why no single similarity measure provides a complete description of the alignment. Adding small isotropic noise to otherwise identical representations leaves CKA near one but reduces local alignment to approximately $0.70$. Removing principal components produces the opposite movement in the two measures. The shared structure is therefore neither a single dominant axis nor a generic equality between representation spaces. Rather, it is a distributed relationship among objects that can survive substantial compression while remaining sensitive to the local arrangement of those objects.

\section{Limitations}

We study fixed pretrained representations rather than models trained explicitly for cross-survey transfer. Our scientific-transfer conclusions concern unchanged linear ridge probes for five catalog targets; they do not rule out nonlinear decodability, a survey-specific adapter, or transfer of every possible physical quantity. A nonlinear adapter, multimodal objective, or joint training procedure could improve transfer, but that would address a different question from the one studied here: whether independently pretrained representations already share physical information in an immediately transferable linear form.

Our experiments also cover many model families within a single scientific domain, with HSC serving as the common reference across the survey comparisons. The physical labels are themselves inferred or measured quantities and therefore carry uncertainties and survey-dependent systematics. The smaller JWST dataset also provides less statistical power than the larger image datasets. Finally, architecture, training objective, model scale, and pretraining data are correlated across the available models. The comparisons in E6 therefore describe associations within this sample rather than showing that any particular model property causes alignment.

\subsection{Compute Cost}

Per-stage embedding extraction across the JWST, HSC, Legacy Survey, and DESI datasets accounted for the majority of the computational cost, particularly when using the larger vision-language models such as PaliGemma and LLaVA. The experiments were conducted across multiple computing environments, including shared GPU infrastructure, a multi-GPU compute node equipped with eight NVIDIA A100 GPUs, 128 CPU cores, and 1~TB of RAM, as well as a local workstation. In total, we estimate that the experiments required approximately 1,500 GPU-hours. Once the embeddings were cached, downstream analyses, including MKNN/CKA computation, linear probing, and statistical analyses, were substantially less computationally intensive, generally requiring approximately 5 GPU-hours per probe on average. Wherever possible, we made the effort to move our computations to GPUs while maintaining mathematical accuracy. 

\begin{table}[t]
\centering
\caption{Summary of computing resources used in the experiments.}
\label{tab:hardware_usage}
\begin{tabular}{ll}
\hline
\textbf{Computing Resource} & \textbf{Usage} \\
\hline
Shared GPU infrastructure & GPU-based embedding extraction and analysis \\
8$\times$ NVIDIA A100 node & GPU-based embedding extraction and analysis \\
128 CPU cores / 1~TB RAM & CPU-intensive downstream analyses \\
Local workstation & Supplementary experiments and analysis \\
\hline
\textbf{Estimated total GPU usage} & $\approx 1{,}500$ GPU-hours \\
\hline
\end{tabular}
\end{table}

\section{NeurIPS Responsibility Statement}
\section*{NeurIPS Paper Checklist}

\begin{enumerate}

\item {\bf Claims}
    \item[] Question: Do the main claims made in the abstract and introduction accurately reflect the paper's contributions and scope?
    \item[] Answer: \answerYes{} % Replace by \answerYes{}, \answerNo{}, or \answerNA{}.
    \item[] Justification: We have clearly stated what we investigate in the abstract, how we do it, what data, and a brief overview of the results and our conclusion is present in both the abstract and the conclusion.

\item {\bf Limitations}
    \item[] Question: Does the paper discuss the limitations of the work performed by the authors?
    \item[] Answer: \answerYes{} % Replace by \answerYes{}, \answerNo{}, or \answerNA{}.
    \item[] Justification: We clearly discuss our limitations in Section C of the appendix. We note how our experiments are limited to model families within a single scientific domain and how our labels themselves carry uncertainties. 

\item {\bf Theory assumptions and proofs}
    \item[] Question: For each theoretical result, does the paper provide the full set of assumptions and a complete (and correct) proof?
    \item[] Answer: \answerNA{} % Replace by \answerYes{}, \answerNo{}, or \answerNA{}.
    \item[] Justification: We do not derive any theoretical results in the paper. 

    \item {\bf Experimental result reproducibility}
    \item[] Question: Does the paper fully disclose all the information needed to reproduce the main experimental results of the paper to the extent that it affects the main claims and/or conclusions of the paper (regardless of whether the code and data are provided or not)?
    \item[] Answer: \answerYes{} % Replace by \answerYes{}, \answerNo{}, or \answerNA{}.
    \item[] Justification: We have calculated four metrics and accounted for confounders such as model size, dimensionality, and biased estimation. This is in the main text of the paper. 

\item {\bf Open access to data and code}
    \item[] Question: Does the paper provide open access to the data and code, with sufficient instructions to faithfully reproduce the main experimental results, as described in supplemental material?
    \item[] Answer: \answerYes{} % Replace by \answerYes{}, \answerNo{}, or \answerNA{}.
    \item[] Justification: In a footnote on the first page, we provide the link to a repository that allows anyone to reproduce our figures. We also provide optional scripts that allow one to fully regenerate our data from upstream datasets, while also noting that such a task requires access to strong multi-GPU machines. 

\item {\bf Experimental setting/details}
    \item[] Question: Does the paper specify all the training and test details (e.g., data splits, hyperparameters, how they were chosen, type of optimizer) necessary to understand the results?
    \item[] Answer: \answerNA{} % Replace by \answerYes{}, \answerNo{}, or \answerNA{}.
    \item[] Justification: Our results rely on frozen pretrained models that are publicly available. 

\item {\bf Experiment statistical significance}
    \item[] Question: Does the paper report error bars suitably and correctly defined or other appropriate information about the statistical significance of the experiments?
    \item[] Answer: \answerYes{} % Replace by \answerYes{}, \answerNo{}, or \answerNA{}.
    \item[] Justification: Figure 1 shows MKNN ranges in lighter shades and it is explicitly addressed in the caption. Figure 3 also includes error bars. Table 4 in the appendix includes ranges of saved-stage difference as well as quantiles. 

\item {\bf Experiments compute resources}
    \item[] Question: For each experiment, does the paper provide sufficient information on the computer resources (type of compute workers, memory, time of execution) needed to reproduce the experiments?
    \item[] Answer: \answerYes{} % Replace by \answerYes{}, \answerNo{}, or \answerNA{}.
    \item[] Justification: In subsection C.1, we provide an estimate of how many GPU-hours it took to complete all of the work. 
    
\item {\bf Code of ethics}
    \item[] Question: Does the research conducted in the paper conform, in every respect, with the NeurIPS Code of Ethics \url{https://neurips.cc/public/EthicsGuidelines}?
    \item[] Answer: \answerYes{} % Replace by \answerYes{}, \answerNo{}, or \answerNA{}.
    \item[] Justification: Our work does not present any anticipated adverse societal or harmful consequences. All experiments were conducted using publicly available foundation models, none of which were trained from scratch as part of this work. The datasets used for embedding extraction consist of astronomical observations focused primarily on galaxies.

\item {\bf Broader impacts}
    \item[] Question: Does the paper discuss both potential positive societal impacts and negative societal impacts of the work performed?
    \item[] Answer: \answerNA{} % Replace by \answerYes{}, \answerNo{}, or \answerNA{}.
    \item[] Justification: Our work focuses primarily on the study of galaxies and astronomical observations, with no direct societal or harmful implications
    
\item {\bf Safeguards}
    \item[] Question: Does the paper describe safeguards that have been put in place for responsible release of data or models that have a high risk for misuse (e.g., pre-trained language models, image generators, or scraped datasets)?
    \item[] Answer: \answerNA{} % Replace by \answerYes{}, \answerNo{}, or \answerNA{}.
    \item[] Justification: 

\item {\bf Licenses for existing assets}
    \item[] Question: Are the creators or original owners of assets (e.g., code, data, models), used in the paper, properly credited and are the license and terms of use explicitly mentioned and properly respected?
    \item[] Answer: \answerYes{} % Replace by \answerYes{}, \answerNo{}, or \answerNA{}.
    \item[] Justification: We have properly credited the MMU collaboration, as well as all creators of the foundation models we used.. 

\item {\bf New assets}
    \item[] Question: Are new assets introduced in the paper well documented and is the documentation provided alongside the assets?
    \item[] Answer: \answerNA{} % Replace by \answerYes{}, \answerNo{}, or \answerNA{}.
    \item[] Justification: We provide the means for readers to reproduce the data product and our figures but do not introduce any new assets. 

\item {\bf Crowdsourcing and research with human subjects}
    \item[] Question: For crowdsourcing experiments and research with human subjects, does the paper include the full text of instructions given to participants and screenshots, if applicable, as well as details about compensation (if any)? 
    \item[] Answer: \answerNA{} % Replace by \answerYes{}, \answerNo{}, or \answerNA{}.
    \item[] Justification: 

\item {\bf Institutional review board (IRB) approvals or equivalent for research with human subjects}
    \item[] Question: Does the paper describe potential risks incurred by study participants, whether such risks were disclosed to the subjects, and whether Institutional Review Board (IRB) approvals (or an equivalent approval/review based on the requirements of your country or institution) were obtained?
    \item[] Answer: \answerNA{} % Replace by \answerYes{}, \answerNo{}, or \answerNA{}.
    \item[] Justification: 

\item {\bf Declaration of LLM usage}
    \item[] Question: Does the paper describe the usage of LLMs if it is an important, original, or non-standard component of the core methods in this research? Note that if the LLM is used only for writing, editing, or formatting purposes and does \emph{not} impact the core methodology, scientific rigor, or originality of the research, declaration is not required.
    %this research? 
    \item[] Answer: \answerNA{} % Replace by \answerYes{}, \answerNo{}, or \answerNA{}.
    \item[] Justification: We do not study any LLMs in this work.

\end{enumerate}

\end{document}